\documentclass[final,1p,times]{elsarticle}

\usepackage{subcaption}

\journal{Computer Physics Communications}

\usepackage{graphicx}
\usepackage{amsmath,amssymb}
\iffalse
\usepackage{booktabs} 
\else
\def\toprule{\hline}
\def\midrule{\hline}
\def\bottomrule{\hline}
\fi
\usepackage{pmbb-sym}
\usepackage{url}
\usepackage[]{units}
\usepackage{color}
\usepackage[dvipsnames]{xcolor}
\usepackage{todonotes}
\usepackage{ragged2e}
\usepackage{algorithm}
\usepackage{algorithmic}

\usepackage{enumitem}

\usepackage{pgfplots} 

\iftrue 

\usepgfplotslibrary{colorbrewer} 
\iftrue
\pgfplotsset{%
  cycle list/Set1, 
  cycle multiindex* list={ 
  mark list*\nextlist
  Set1\nextlist},                    
}                           
\fi

\fi

\usepackage{ulem}

\def\fat#1{\boldsymbol{#1}}
\def\norm#1{\lVert{#1}\rVert}
\def\abs#1{\left|{#1}\right|}

\definecolor{darkgreen}{cmyk}{1,0,1,0.608}
\definecolor{darkorange}{rgb}{200,51,0}

\def\red#1{\textcolor{red}{#1}}

\def\navyblue#1{\textcolor{NavyBlue}{#1}}
\def\blue#1{\navyblue{#1}}

\def\orange#1{\textcolor{BurntOrange}{#1}}

\def\XXint#1#2#3{{\setbox0=\hbox{$#1{#2#3}{\int}$}\vcenter{\hbox{$#2#3$}}\kern-.5\wd0}}

\def\NL{\mathcal{N}}
\def\IL{\mathcal{I}}
\def\CL{\mathcal{C}}

\def\separation{\sigma} 

\newcommand{\dist}{\mathop{\mathrm{dist}}}

\usepackage{amsthm}
\newtheorem{remark}{Remark}

\def\lFMM{\ell_{\rm FMM}}
\def\lBD{\ell_{\rm BD}}

\def\PtoP{P2P}
\def\PtoM{P2M}
\def\MtoM{M2M}
\def\MtoL{M2L}
\def\LtoL{L2L}
\def\LtoP{L2P}
\def\PtoL{P2L} 
\def\MtoP{M2P} 

\def\appendixname{}

\iffalse
\usepackage{numprint}
\npthousandsep{} 
\npdecimalsign{.}
\npproductsign{\times}
\npaddplusexponent
\npaddmissingzero
\else
\usepackage{siunitx}
\NewDocumentCommand\numprint{m}{\num[round-mode = places]{#1}}
\NewDocumentCommand\nprounddigits{m}{\sisetup{round-precision = #1}}
\def\npproductsign#1{} 
\fi

\newcolumntype{H}{>{\setbox0=\hbox\bgroup}c<{\egroup}@{}} 

\nprounddigits{1} 
\def\prtime#1{{\nprounddigits{0}\textrm{\numprint{#1}}}} 
\def\prval#1{{\nprounddigits{1}\textrm{\numprint{#1}}}} 
\iffalse
\def\prspeedup#1{{\nprounddigits{1}\textbf{\numprint{#1}}}} 
\def\prniterspeedup#1{{\nprounddigits{1}\textbf{\numprint{#1}}}} 
\else
\def\prspeedup#1{{\nprounddigits{1}\textbf{\sisetup{detect-weight}\numprint{#1}}}} 
\def\prniterspeedup#1{{\nprounddigits{1}\textbf{\sisetup{detect-weight}\numprint{#1}}}} 
\fi
\def\prerr#1{{\nprounddigits{1}\npproductsign{\times}\textrm{\numprint{#1}}}} 
\def\prmaxerr#1{{\nprounddigits{1}\npproductsign{\times}\sisetup{output-exponent-marker=\text{e}}\textrm{\numprint{#1}}}} 

\def\noteIV{}
\def\noteVIII{}
\def\noteXVI{}
\def\noteXXXII{}
\def\noteLXIV{} 

\newcolumntype{M}{p{35pt} >{\RaggedLeft\arraybackslash}p{10pt} >{\RaggedLeft\arraybackslash}p{15pt} H >{\RaggedLeft\arraybackslash}p{22pt} >{\RaggedLeft\arraybackslash}p{13pt} >{\RaggedLeft\arraybackslash}p{10pt} >{\RaggedLeft\arraybackslash}p{18pt} >{\RaggedLeft\arraybackslash}p{18pt} >{\RaggedLeft\arraybackslash}p{20pt} >{\RaggedLeft\arraybackslash}p{15pt} >{\RaggedLeft\arraybackslash}p{15pt} H H H >{\RaggedLeft\arraybackslash}p{20pt} H} 

\def\myunderlineIV{}
\def\myunderlineVIII{}
\def\myunderlineXVI{}
\def\myunderlineXXXII{}
\def\myunderlineLXIV{} 

\newcommand{\niter}{niter} 
\newcommand{\nsp}{nsp}     
\newcommand{\main}{total}  
\newcommand{\tsp}{sp}      
\newcommand{\ptop}{p2p}    
\newcommand{\gmres}{gmres} 
\newcommand{\mv}{mv}       
\newcommand{\ms}{prec}     
\newcommand{\init}{init}   
\newcommand{\elim}{fact}   
\newcommand{\umin}{umin}   
\newcommand{\umax}{umax}   
\newcommand{\llerr}{l2err} 
\newcommand{\merr}{maxerr}   
\newcommand{\Note}{Note}   

\gdef\myheader#1{\if1#1 \toprule Precon. & $k$ & \niter & \nsp & \main & \tsp & \ptop & \gmres & \mv & \ms & \init & \elim & \umin & \umax & \llerr & \merr & \Note\\ \midrule \else \midrule \fi} 
\gdef\myfooter#1{\if1#1 \bottomrule \else \fi}

\def\myheaderFlag{}
\def\myfooterFlag{}

\def\del#1{}
\usepackage{framed} 

\def\autoref#1{\ref{#1}}
\def\equationautorefname~#1\null{%
  Eq.~(#1)\null
}
\def\algorithmautorefname~#1\null{%
  Algorithm~#1\null
}
\def\sectionautorefname~#1\null{%
  Section~#1\null
}
\def\subsectionautorefname~#1\null{%
  Section~#1\null
}
\def\subsubsectionautorefname~#1\null{%
  Section~#1\null
}
\def\tableautorefname~#1\null{%
  Table~#1\null
}
\def\figureautorefname~#1\null{%
  Figure~#1\null
}
\def\lineautorefname~#1\null{%
  Line~#1\null
}

\begin{document}

\begin{frontmatter}

\title{Parallelization of the inverse fast multipole method with an application to boundary element method}

  
\author[NU]{Toru Takahashi\corref{cor}}
\ead{toru.takahashi@mae.nagoya-u.ac.jp}
\cortext[cor]{Corresponding author}
\author[SU1]{Chao Chen}
\author[SU1,SU2]{Eric Darve}

\address[NU]{Department of Mechanical Systems Engineering, Nagoya University, Furo-cho, Nagoya, Aichi, 464-8603 Japan}
\address[SU1]{Institute for Computational \& Mathematical Engineering, Stanford University, 496 Lomita Mall, 94305, Stanford, CA, USA}
\address[SU2]{Department of Mechanical Engineering, Stanford University, 496 Lomita Mall, 94305, Stanford, CA, USA}






\begin{abstract}

We present an algorithm to parallelize the inverse fast multipole method (IFMM), which is an approximate direct solver for dense linear systems. The parallel scheme is based on a greedy coloring algorithm, where two nodes in the hierarchy with the same color are separated by at least $\separation$ nodes. We proved that when $\separation \ge 6$, the workload associated with one color is embarrassingly parallel. However, the number of nodes in a group (color) may be small when $\separation = 6$. Therefore, we also explored $\separation = 3$, where a small fraction of the algorithm needs to be serialized, and the overall parallel efficiency was improved. We implemented the parallel IFMM using OpenMP for shared-memory machines. Successively, we applied it to a fast-multipole accelerated boundary element method (FMBEM) as a preconditioner, and compared its efficiency with (a) the original IFMM parallelized by linking a multi-threaded linear algebra library and (b) the commonly used parallel block-diagonal preconditioner. Our results showed that our parallel IFMM achieved at most $4\times$ and $11\times$ speedups over the reference method (a) and (b), respectively, in realistic examples involving more than one million variables.
\end{abstract}


\end{frontmatter}

\section{Introduction}\label{s:intro}

We are interested in solving a large-scale dense linear system 
\begin{eqnarray}
  Ax = b
  \label{eqn:axb}
\end{eqnarray}
with $N$ unknown variables. Such systems arise from many science and engineering areas, such as discretized integral equations, boundary element methods (BEMs), machine learning, etc. To solve \autoref{eqn:axb}, a naive Gaussian elimination requires $O(N^3)$ computation and $O(N^2)$ memory, which is prohibitive when $N$ is large. While an iterative solver such as CG and GMRES requires $O(N^2)$ computation and memory per iteration, the number of iterations can be large in presence of ill-conditioning and indefinite problems and the $O(N^2)$ costs limit their application to truly large-scale problems. To overcome these disadvantages, the fast multipole method (FMM)~\cite{1985_Rokhlin,1987_Greengard,Darve2000195,2002_Nishimura} can be used to accelerate the computation of matrix-vector products at every iteration of an iterative method, reducing the computation and memory costs to typically $O(N)$.


In order to further reduce the number of iterations of an iterative method, the inverse fast multipole method (IFMM)~\cite{2014_Ambikasaran,coulier2017} has been developed among various fast direct solvers~\cite{2005_Martinsson,2015_Corona,2015_Bremer,2016_Ho} and $\mathcal{H}$-matrix methods~\cite{bebendorf2008hierarchical, 2008_Banjai}. These methods perform Gaussian elimination on a compressed representation of $A$ and computes an approximated factorization of $A$ with nearly linear computational cost. The IFMM is inspired and closely related to the FMM since they use the same hierarchical decomposition of the problem domain (a.k.a. the FMM tree), which corresponds to a partitioning of the unknowns in \autoref{eqn:axb}, and use the same far-field (multipole-to-local) and near-field (particle-to-particle) translation operators, which corresponds to the full-rank and the low-rank matrix blocks in $A$, respectively. This allows the IFMM to solve \autoref{eqn:axb} with the same asymptotic computation and memory complexities as the FMM, i.e., $O(N\log^2\frac{1}{\varepsilon})$, where $\varepsilon$ is a prescribed accuracy. The efficiency of the IFMM has been studied in several previous works for solving \autoref{eqn:axb}: Quaife et al~\cite{quaife2017} and Coulier et al~\cite{coulier2017} applied the IFMM as preconditioners for the GMRES~\cite{saad1986gmres} to solve \autoref{eqn:axb} from the immersed boundary method regarding Stokes flow problems in two and three dimensions (3D); Takahashi et al~\cite{takahashi2017} applied the IFMM together with the low-frequency FMM to accelerate the BEM for the 3D Helmholtz equation; Coulier et al~\cite{ij-cmame-coul-16a} applied the IFMM to reduce the cost of a mesh deformation method which is based on the radial basis function interpolation.

To further reduce the runtime of the IFMM for large problem sizes ($\sim1$ million variables), this paper introduces the parallel IFMM algorithm targeting at shared-memory machines. Although the IFMM uses the same hierarchical tree decomposition as the FMM, the data dependency of the IFMM is much more complicated: the work associated with nodes at the same level in the hierarchy is \textit{not} embarrassingly parallel as it is in the upward pass and the downward pass of the FMM~\cite{2011_Darve,yokota2011biomolecular}. Rather than the FMM, the data dependency in the IFMM is similar to that in the incomplete Cholesky factorization~\cite{saad1996iterative} but has a special structure associated with the FMM hierarchical decomposition. Rather than solving \autoref{eqn:axb} directly, the IFMM solves the equivalent \textit{extended sparse linear system}~\cite{coulier2017}, which is associated with the FMM tree structure. As defined in the original FMM, every node in the FMM tree has a list of neighbors and an interaction list of nodes in the far field. In the IFMM, one step of (block) Gaussian elimination on a node in the FMM tree leads to a dense Schur complement among its neighbors (a.k.a., clique). The Schur complement is then compressed based on the low-rank structure in the FMM, which effectively updates the far-field translation operators of the neighbors and affects the nodes in the neighbors' interaction lists.

Our parallel IFMM algorithm is based on a greedy coloring algorithm (\autoref{algo:group}), which assigns the same color to nodes that are apart from each other by a distance of $\separation$. We proved that the workload for nodes in the same group (or nodes with the same color) is embarrassingly parallel when $\separation \ge 6$, or technically speaking, the neighbors' interaction lists of a node $n_1$ do not intersect with the neighbors of another node $n_2$ if $n_1$ and $n_2$ have the same color. So we derived the basic parallel IFMM algorithm (\autoref{s:full}) associated with $\separation = 6$. For the targeting problem sizes ($\sim1$ million), we found that the number of colors is large (the average group size is small), which limits the parallelism and leads to a significant synchronization cost.

Therefore, we derived the improved parallel IFMM algorithm (\autoref{s:part}) associated with $\separation = 3$ in our coloring algorithm. With $\separation = 3$, there exists a race condition for processing two nodes with the same color: their neighbors may need to update the same far-field translation operator. So these updates are serialized and form a critical section in our improved parallel algorithm. Fortunately, these far-field updates account for only a small fraction of the entire computation (see Fig. 26 in~\cite{coulier2017}), and the overhead of serialization is paid back by the tremendous improvement of parallel efficiency from large group sizes and a small number of groups/colors.

In our previous study~\cite{takahashi2017}, a naive parallel IFMM algorithm was developed by simply replacing all matrix computations such as matrix-matrix multiplication, LU decomposition, and the SVD with their parallel counterparts in the sequential IFMM algorithm. This approach can be realized easily by linking the sequential IFMM code with a multi-threaded linear algebra library such as the Intel MKL library~\cite{wang2014intel} and the OpenBLAS library~\cite{OpenBLAS}. However, this over-fine-grained parallelization would inevitably introduce overwhelming overhead when the number of threads is large. Another closely related approach is the distributed-memory parallel solver introduced by Chen et al~\cite{chen2018}, which parallelizes the LoRaSp algorithm~\cite{pouransari2017}, an analog of the IFMM for solving \autoref{eqn:axb} when $A$ is sparse. 

To summarize, this paper presents parallel IFMM algorithms for solving large-scale dense linear systems, and especially, our work makes the following three main contributions:
\begin{enumerate}
\item
analyzing the data dependency and identifying the parallelism in the sequential IFMM algorithm.
\item
deriving two parallel IFMM algorithms (\autoref{s:full} and \autoref{s:part}) based on our coloring scheme (\autoref{algo:group}).
\item
reporting benchmarks and comparisons between our parallel IFMM algorithms and two existing methods for solving real-world problems involving more than a million variables.
\end{enumerate}

The rest of the paper is organized as follows. \autoref{s:ifmm} reviews the basics of the IFMM algorithm. \autoref{s:parallel} describes our parallel strategy and an implementation using OpenMP~\cite{OpenMP}. \autoref{s:preconditioner} incorporates the parallel IFMM to solve dense linear systems from the fast-multipole accelerated BEM (FMBEM) for 3D Helmholtz equation. \autoref{s:num} gives two numerical examples to assess the efficiency of our parallel algorithms. \autoref{s:concl} concludes this study.




\section{IFMM}\label{s:ifmm}

In this section, we describe the sequential IFMM algorithm focusing on the data dependency there and introduce some FMM terminologies used in this paper. For more details, the readers may see the references~\cite{2014_Ambikasaran,coulier2017}.

IFMM is a fast direct solver with $O(N\log^2\frac{1}{\varepsilon})$ complexity for solving a linear system $Ax=b$, where the matrix-vector product with $A \in \bbbr^{N \times N}$ can be evaluated with the FMM. This type of linear system arises from various science and engineering applications. As an example, consider an electrostatic field generated by $N$ charges, where the $i$-th charge is located at $\fat{p}_i\in\bbbr^3$ and the electric potential is given as $b_i$ at $\fat{p}_i$. We are interested in solving for the charge density $x_i$. This problem can be formulated as $Ax=b$, where $x:=[x_1,\ldots,x_N]^T$, $b:=[b_1,\ldots,b_N]^T$, and the $(i,j)$-th element of $A$ is given by the Coulomb interaction, i.e., $\frac{x_i}{4\pi\epsilon\abs{\fat{p}_i-\fat{p}_j}}$ if $i\ne j$ and $0$ otherwise, where $\epsilon$ denotes a permittivity. To solve such a linear system, IFMM takes the following four steps.

\subsection{Hierarchical domain decomposition}\label{s:hdd}

We carry out the standard hierarchical partitioning of the problem domain as in the FMM~\cite{1987_Greengard,Darve2000195,2002_Nishimura} and associate the partitioning with an \textit{octree} (\textit{quadtree}) in a three (two) dimensional space. This partitioning can be easily parallelized with the geometric information of the problem domain. For ease of presentation, we assume the tree is \textit{uniform} (a.k.a., \textit{complete}), so there are $n_{\kappa} = 2^{\kappa \, d}$ nodes at level $\kappa = 0, 1, \ldots, \ell$, where $d$ is the space dimension. Based on the partitioning, we also define the \textit{neighbors} and the \textit{interaction list} of a tree node $i$ as in the FMM (a partition refers to a (cubic) cluster of points in the $d$-dimensional space): 
\begin{itemize}
\item[]
\textbf{Neighbors}: tree nodes corresponding to adjacent partitions of node $i$ (excluding $i$);
\item[]
\textbf{Interaction list}: non-adjacent partitions whose parent is a neighbor of node $i$'s parent.
\end{itemize}

\begin{remark}{}
Our definition of neighbors obeys the standard definition in graph theory but differs from one in the FMM, where the neighbors of a node includes the node itself.
\end{remark}

\subsection{Initialization of tree data structure}\label{s:init_tree_data}

   Given that the FMM can be used to compute matrix-vector product with $A$ in \autoref{eqn:axb}, we initialize our tree data structure with the six translation operators in the FMM: P2P, P2M, L2P, M2M, M2L, and L2P. In particular, every leaf node stores all six FMM operators associated its partition, and every internal tree node stores M2M, M2L, and L2P associated with its partition (the other three operators are initially empty and will be computed in the algorithm; see lines \autoref{line:fact_transfer1} -- \autoref{line:fact_transfer3} in \autoref{algo:fact}). In the context of BEM, the expressions of the FMM translation operators can be found in~\cite{takahashi2017}. Note this initialization step is embarrassingly parallel with respect to all tree nodes, and this was introduced in our previous work~\cite{takahashi2017}.
  
\begin{remark}{}
   From an algebraic perspective, our tree data structure corresponds to the graph of the extended sparse linear system~\cite{coulier2017} of \autoref{eqn:axb} (see Figure 10 in \cite{coulier2017}). Specifically, the extended sparse linear system is constructed through introducing auxiliary variables into \autoref{eqn:axb}, which are nothing but the ``multipole moments'' and the ``local coefficients'' in the FMM. These auxiliary variables satisfy relations as defined in the FMM, and solving the extended system is equivalent to solving \autoref{eqn:axb}. As an example, assume a three-level hierarchy in the domain decomposition, the extended sparse linear system is the following:
  
\begin{eqnarray}
  \begin{bmatrix}
    \PtoP^{(3)} & \LtoL^{(3)} & & &\\
    {\MtoM^{(3)}}^{\mathrm{T}} & & -I	& &\\ 
    & -I & \MtoL^{(3)} & \LtoL^{(2)} &\\
    & & {\MtoM^{(2)}}^{\mathrm{T}} & & -I\\
    & & & -I & \MtoL^{(2)}
  \end{bmatrix}
  \begin{Bmatrix}
    x^{(3)}\\
    z^{(3)}\\
    y^{(3)}\\
    z^{(2)}\\
    y^{(2)}
  \end{Bmatrix}
  =
  \begin{Bmatrix}
    b\\
    0\\
    0\\
    0\\
    0
  \end{Bmatrix},
  \label{eq:ext}
\end{eqnarray}
where $x^{(3)}$ are exactly the unknowns in \autoref{eqn:axb}, $y^{(\kappa)}$ and $z^{(\kappa)}$ are the multipole moments and the local coefficients associated with all the tree nodes at level $\kappa$, $I$ denotes the identity matrix, and the other matrix blocks correspond to FMM translation operators at different levels.
\end{remark}

\subsection{Factorization}

After initializing our tree data structure, we have effectively transformed the original $N \times N$ dense problem to a (block) sparse linear system with $O(N)$ nonzero entries. To compute an approximate factorization of the sparse matrix in linear complexity, the IFMM employs both Gaussian elimination and low-rank approximation of fill-in blocks.

The factorization phase of the IFMM is summarized in \autoref{algo:fact}, which is a level-by-level traversal of the tree from the leaf level towards the root, as the upward pass in the FMM. For tree nodes at the same level, an elimination-and-compression strategy is conducted iteratively as follows.

At a node $i$, (block) Gaussian elimination is carried out on the associated variables (line \autoref{line:fact_eliminate}), leading to fill-in blocks among its neighboring nodes (a.k.a., \textit{clique}). Then, a compression step is carried out on all pairs of nodes that are neighbors of $i$ (line \autoref{line:fact_pairs}). For such a pair of nodes $(p,q)$, the IFMM may compute a low-rank approximation of the associated fill-in block and update existing FMM operators, depending on whether $p$ or $q$ has been visited/eliminated and whether they are in each other's interaction list (lines \autoref{line:fact_if_zero}, \autoref{line:fact_if_one}, \autoref{line:fact_if_two} and \autoref{line:fact_if_three}). In particular, if a node $q$ is not eliminated and $p$ is in $q$'s interaction list, node $q$ needs to update the stored M2L operators corresponding to its interaction list (lines \autoref{line:fact_YN} and \autoref{line:fact_NN2}). The same principle applies to node $p$ (line \autoref{line:fact_NN1}). Finally, after level $\kappa$ is traversed, the IFMM aggregates the M2M, L2L, and M2L operators of nodes that have the same parent to form the P2M, L2P, and P2P operators of their parent at level $\kappa - 1$ for $\kappa > 2$ (lines \autoref{line:fact_transfer1} and \autoref{line:fact_transfer3}). When $\kappa = 2$, a dense linear system is formed with all M2L operators of the level and factorized with the standard LU factorization; this step is intentionally omitted in \autoref{algo:fact} to save space. 

As the above shows, the elimination-and-compression process on a tree node involves only its neighbors and their interaction lists. So two sufficiently distant nodes at the same level can be processed in parallel, which forms the basis of our parallel algorithm in \autoref{s:parallel}.

\begin{remark}{}
Algebraically speaking, we have computed an implicit factorization of the extended sparse linear system (with $O(N)$ nonzeros), which consists of (block) triangular factors and (block) diagonal factors.
\end{remark}

\subsection{Solve}

With the (implicit) factorization, the solve phase follows the standard (sparse) forward and backward substitutions. As \autoref{algo:solve} shows, the in-place solve algorithm consists of three parts: initialization, upward traversal, and downward traversal of our tree data structure. For the initialization, every leaf node extracts corresponding elements in a given right-hand size, and every internal tree node gets a zero vector. It is obvious that this initialization can be easily parallelized among all nodes in the tree.

Both forward and backward substitutions employ the level-by-level traversal as in the factorization, and local updates on the solution vector of node $i$ involves only the neighbors of $i$ (lines \autoref{line:n1} and \autoref{line:n2}). Assume two nodes $i$ and $j$ can be processed in parallel during the factorization, then it is easy to see that they can also be processed in parallel during the two substitutions.

\begin{remark}{}
As in the factorization, we omit the solve with respect to the (small) linear system at level 1, which is done with standard (dense) forward and backward substitutions. 
\end{remark}

\begin{remark}{}
\autoref{algo:solve} is a simplification of the algorithm presented in~\cite{2014_Ambikasaran,coulier2017}, where the original solve algorithm involves not only $x^{(\kappa)}$ but also multipole moments $y^{(\kappa)}$ and local coefficients $z^{(\kappa)}$. The data dependency in the solve algorithm is exactly the same.
\end{remark}


\begin{algorithm}[h]
  \caption[Caption for IFMM factorization]{Sequential factorization algorithm. Notations: $\NL_i$ and $\IL_i$ stand for the \textit{neighbors} and the \textit{interaction list} of node $i$, respectively; $\tilde{p}, \tilde{q}$ denote the parents of node $p$ and node $q$, respectively; an FMM operator is denoted as $A_{pq}$, where $A$ represents the operator name, and $p$ and $q$ stand for the \textit{target} and \textit{source} node~\cite{2014_Ambikasaran}, respectively. Underlined are the critical sections (in \autoref{s:part}), and comments in \orange{orange} show the data dependencies (a)--(c) in \autoref{s:race}.}
  \label{algo:fact}
  \small
  \begin{algorithmic}[1]
    \FOR{Level $\kappa=\ell$ \TO $2$} \label{line:fact_loop_kappa}    
    \FOR{Node $i=0$ \TO $n_\kappa-1$} \label{line:fact_loop_i}

    \STATE Factorize $P2P_{ii}$ and ``eliminate'' node $i$ \orange{// (a); local and neighbors of $i$}  \label{line:fact_eliminate}
        
    \FORALL{Pair $(p,q)$, where $p,q\in\NL_i$} \label{line:fact_pairs}
    
    \IF{Node $p$ is eliminated \textbf{\&} node $q$ is eliminated} \label{line:fact_if_zero}
    \STATE  \blue{// Case 1}

    \STATE Calculate and add fill-in blocks to $\MtoL_{pq}$ and $\MtoL_{qp}$ \orange{// (b); local of $p$ and $q$} \label{line:fact_case0_M2L}

    \ELSIF{Node $p$ is eliminated \textbf{\&} node $q$ is not eliminated} \label{line:fact_if_one}
    \STATE \blue{// Case 2}
    
    \STATE Calculate fill-in blocks $\PtoL_{pq}^\prime$ and $\MtoP_{qp}^\prime$ \orange{// (b); local of $p$ and $q$} \label{line:fact_case1_first}

    \IF{Node $p\in\IL_q$} \label{line:fact_case1_second}
    \STATE Compress fill-in $\PtoL_{pq}^\prime \approx \MtoL_{pq}^\prime \PtoM_{qq}^\prime$ \orange{// (b); local of $q$}  \label{line:fact_case1_nn_start}
    \STATE Compress fill-in $\MtoP_{qp}^\prime \approx \LtoP_{qq}^\prime \MtoL_{qp}^\prime$ \orange{// (b); local of $p$} 
    
    \STATE Update $\PtoM_{qq}$ and $\LtoP_{qq}$ \orange{// (b); local of $q$} \label{line:fact_case1_nn_end}
    \STATE \uline{Update $\MtoL_{lq}$ and $\MtoL_{ql}$ for $l\in\IL_p$} \orange{// local and interaction list of $q$} \label{line:fact_YN} 
    \STATE Update $\MtoM_{\tilde{q}q}$ and $\LtoL_{q\tilde{q}}$ \orange{// (b); local of $q$} \label{line:fact_YN_parent}
    
    \ELSE
    \STATE Add fill-in blocks to $\PtoL_{pq}$ and $\MtoP_{qp}$ \orange{// (b); local of $p$ and $q$}  \label{line:fact_case1_NL} 
    \ENDIF \label{line:fact_case1_last}

    \ELSIF{Node $p$ is not eliminated \textbf{\&} node $q$ is eliminated} \label{line:fact_if_two}
    \STATE \blue{// Case 3: reverse $p$ and $q$ in Case 2 (skipped)}

    \ELSIF{Node $p$ is not eliminated \textbf{\&} node $q$ is not eliminated} \label{line:fact_if_three}
    \STATE \blue{// Case 4}
    \STATE Calculate fill-in blocks $P2P_{pq}^\prime$ and $P2P_{qp}^\prime$ \orange{// (b); local of $p$ and $q$}
    
    \IF{Node $p\in\IL_q$} 
    \STATE Compress fill-in $P2P_{pq}^\prime \approx \LtoP_{pp}^\prime \MtoL_{pq}^\prime \PtoM_{qq}^\prime$ \orange{// (b); local of $q$}
    \STATE Compress fill-in $P2P_{qp}^\prime \approx \LtoP_{qq}^\prime \MtoL_{qp}^\prime \PtoM_{pp}^\prime$ \orange{// (b); local of $p$}
    \STATE Update $\PtoM_{pp}$, $\LtoP_{pp}$, $\PtoM_{qq}$, and $\LtoP_{qq}$ \orange{// (b); local of $p$ and local of $q$}
    \STATE \uline{Update $\MtoL_{lp}$ and $\MtoL_{pl}$ for $l\in\IL_p$} \orange{// (b) and (c); local and interaction list of $p$}  \label{line:fact_NN1} 
    \STATE \uline{Update $\MtoL_{lq}$ and $\MtoL_{ql}$ for $l\in\IL_q$} \orange{// (b) and (c); local and interaction list of $q$}  \label{line:fact_NN2} 
    \STATE Update $\MtoM_{\tilde{p}p}$ and $\LtoL_{p\tilde{p}}$ \orange{// (b); local of $p$} \label{line:fact_NN1_parent}
    \STATE Update $\MtoM_{\tilde{q}q}$ and $\LtoL_{q\tilde{q}}$ \orange{// (b); local of $q$} \label{line:fact_NN2_parent}
    \ELSE
    \STATE Add fill-in blocks to $\PtoP_{pq}$ and $\PtoP_{qp}$ \orange{// (b); local of $p$ and $q$}
    \ENDIF
    \ENDIF

    \ENDFOR 
    \ENDFOR 

    \IF{$\kappa>2$}
    \FOR{Node $i=0$ \TO $n_{\kappa-1}-1$}
    \STATE Aggregate M2M and L2L from children to form P2M and L2P. \orange{// (a); local of $i$} \label{line:fact_transfer1}
    \STATE Aggregate M2L from children to form P2P. \orange{// (a); local and neighbors of $i$} \label{line:fact_transfer3}
    \ENDFOR 
    \ENDIF
    
    \ENDFOR 
  \end{algorithmic}
\end{algorithm}


\begin{algorithm}[h]
  \caption{Sequential (in-place) solve algorithm. Input: right-hand side $b$; output: solution $x$ ($\equiv x^{(\ell)}$). Notations: $\NL_i$ and $\CL_i$ stand for the \textit{neighbors} and \textit{children} of node $i$. Comments in \orange{orange} show the data dependencies (a) and (b) in \autoref{s:race}; there is no dependency of (c).}
  \label{algo:solve}
  \begin{algorithmic}[1]
    \STATE \blue{// Initialize node vector}
    \FOR{Level $\kappa=\ell$ \TO $2$}
    \FOR{Node $i=0$ \TO $n_\kappa-1$} \label{line:b_loop_i}
    \IF{$\kappa==\ell$}
    \STATE $x^{(\kappa)}_i = b_i$ \blue{// corresponding components in the RHS vector $b$}
    \ELSE
    \STATE $x^{(\kappa)}_i = 0$
    \ENDIF
    \ENDFOR
    \ENDFOR

    \STATE \blue{// (column-oriented) forward substitution/upward pass}

    \FOR{Level $\kappa=\ell$ \TO $2$}
    
    \FOR{Node $i=0$ \TO $n_\kappa-1$} \label{line:elimination_loop_i}
    
    \STATE Mark node $i$ as ``visited'' \orange{// (a); local of $i$} \label{line:visit}

    \FORALL{Node $p\in\NL_i$} \label{line:n1}
    
    \IF{Node $p$ is visited} \label{line:solve_if_one}

    \STATE $x_p^{(\kappa)} -= \PtoL_{pi} x_i^{(\kappa)}$ \label{line:elimination_P2L} \orange{// (b); local of $p$} 

    \ELSE 
    
    \STATE $x_p^{(\kappa)} -= \PtoP_{pi} x_i^{(\kappa)}$ \label{line:elimination_P2P} \orange{// (b); local of $p$} 

    \ENDIF

    \ENDFOR

    \ENDFOR
    
    \FOR{Node $i=0$ \TO $n_{\kappa-1} -1$}

    \STATE Aggregate $x_{\CL_i}^{(\kappa)}$ to form $x_{i}^{(\kappa - 1)}$ \orange{// (a); local of $i$} 

    \ENDFOR
    \ENDFOR

           
    \STATE \blue{// (row-oriented) backward substitution/downward pass}
    \FOR{Level $\kappa=2$ \TO $\ell$} \label{line:substitution_loop_level}
    
    \FOR{Node $i=n_\kappa-1$ \TO $0$} \label{line:substitution_loop_i}
    
    \FORALL{Node $p\in\NL_i$}  \label{line:n2}
    

    \IF{Node $p$ is visited} \label{line:solve_if_two}

    \STATE $x_i^{(\kappa)} -= \MtoP_{ip} x_p^{(\kappa)}$ \orange{// (a); local of $i$} \label{line:elimination_M2P}

    \ELSE 
    
    \STATE $x_i^{(\kappa)} -= \PtoP_{ip} x_p^{(\kappa)}$ \orange{// (a); local of $i$} 

    \ENDIF

    \ENDFOR

    \STATE Solve $x_i^{(\kappa)}$ with the local factorization of $\PtoP_{ii}$ \orange{// (a); local of $i$} 

    \STATE Mark node $i$ as ``un-visited'' \orange{// (a); local of $i$} 
    
    \STATE Split $x_i^{(\kappa)}$ into $x^{(\kappa + 1)}_{\CL_i}$ \orange{// (a); local of $\CL_i$} \label{line:substitution_child_mom}

    \ENDFOR

    \ENDFOR \label{line:substitution_loop_level_end}
  \end{algorithmic}
\end{algorithm}

\section{Parallelization of IFMM}\label{s:parallel}

In this section, we describe the parallelization for each of the four steps in the IFMM. The first step --- hierarchical domain decomposition --- can be done using existing parallel partitioning packages, such as Zoltan~\cite{ZoltanIsorropiaOverview2012}. The second step --- initialization of tree data structure --- is embarrassingly parallel with respect to all tree nodes, and this has been introduced in our previous work~\cite{takahashi2017}, so we skip the details here. The last two steps --- factorization and solve --- are the main challenges and the focus of this section.

As hinted earlier, our parallel strategy is based on coloring: assigning colors to tree nodes at the same level such that nodes of the same color can be processed in parallel during the factorization and the solve. To be more specific, our parallel algorithm replaces the loop over all tree nodes at a level (line \autoref{line:fact_loop_i} in \autoref{algo:fact}, and lines \ref{line:elimination_loop_i} and \ref{line:substitution_loop_i} in \autoref{algo:solve}) with nested loops, where the outer loop goes over all colors \textit{sequentially}, and the inner loop goes over tree nodes with the same color \textit{in parallel}. 

To that end, we introduce the separation parameter $\separation$, which characterizes the minimum (normalized) distance between all pairs of nodes having the same color, and also serves as the constraint in our coloring algorithm. Let the partitions be cubes of size $L$ at a level in the FMM-tree, and the normalized (infinity norm) distance between two tree nodes $u$ and $v$ is defined as 
\begin{eqnarray*}
\dist(u,v):=\frac{\max(\abs{x_u-x_v},\abs{y_u-y_v},\abs{z_u-z_v})}{L},
\end{eqnarray*}
where $(x_u,y_u,z_u)$ and $(x_v,y_v,z_v)$ are the coordinates of the two partition (cube) centers associated with $u$ and $v$, respectively. Note that by construction, $\dist(u,v)$ is a non-negative integer, and 
\begin{eqnarray} \label{eq:dist}
\dist(u,v) = 
\begin{cases}
0 & \quad \text{$u=v$,} \\
1 & \quad \text{$u \in \NL_v (\Leftrightarrow v \in \NL_u)$},\\
2 \text{ or } 3 & \quad \text{$u\in \IL_v (\Leftrightarrow v\in \IL_u)$.} 
\end{cases}
\end{eqnarray}
The separation parameter, or coloring constraint is thus defined as 
\begin{eqnarray}
  \separation = \min \, \{ \, \dist(u,v) \, | \, \text{color}(u) = \text{color}(v) \, \},
  \label{eq:sep}
\end{eqnarray}
where the function ``color()'' returns the color (an integer) of a node.

In the following, we analyze the data dependency in \autoref{algo:fact} and \ref{algo:solve} and show what value $\separation$ needs to take in the coloring algorithm.

\subsection{Data dependency}\label{s:race}




Suppose we have two threads $T_i$ and $T_{i'}$, and want to process two nodes $i$ and $i'$ independently. Apparently, $i \not= i'$ means $\separation \ge 1$. The data dependency in the IFMM falls into three cases as follows.

\begin{enumerate}[label=(\alph*)]

\item \textbf{Local operations} 

In \autoref{algo:fact} and \ref{algo:solve}, $T_i$ (resp. $T_{i'}$) updates local data of node $i$ (resp. $i'$) and reads data from neighbors of $i$ (resp. $i'$). For example, $T_i$ updates the ``eliminated'' flag (line \autoref{line:fact_eliminate}) in \autoref{algo:fact} and the ``visited'' flag (line \autoref{line:visit}) in \autoref{algo:solve}, and reads these flags of node $i$'s neighbors (\autoref{algo:fact}: lines \autoref{line:fact_if_zero}, \autoref{line:fact_if_one}, \autoref{line:fact_if_two} and \autoref{line:fact_if_three}, and \autoref{algo:solve}: lines \autoref{line:solve_if_one} and \autoref{line:solve_if_two}). If node $i'$ coincides with $i$'s neighbor $p$, the flags of $i'$ can be updated by $T_{i'}$ while $T_i$ is working on node $i$ according to the original flags of $p (=i')$. This is thus a race condition. To avoid it, $i$ and $i'$ should not be in each other's neighbor list, i.e., $i\notin\NL_{i'}$ ($\Leftrightarrow i'\notin\NL_i$). According to \autoref{eq:dist}, this is guaranteed if
  \begin{eqnarray*}
    \separation \ge 2.
    \label{eq:sep2}
  \end{eqnarray*}


  

\item \textbf{Neighbor operations} 


In \autoref{algo:fact} and \ref{algo:solve}, $T_i$ (resp. $T_{i'}$) updates the local data of $i$'s (resp. $i'$'s) neighbors, i.e., nodes $p$ and $q$ (resp. $p'$ and $q'$). This includes two scenarios as the following. First, $T_i$ updates operators/data associated with a single neighbor of $i$, e.g., $\MtoM_{\tilde{q}q}$ and $\LtoL_{q\tilde{q}}$ (lines \ref{line:fact_YN_parent} and \ref{line:fact_NN2_parent}) in \autoref{algo:fact} and $x_p^{(\kappa)}$ (line \autoref{line:elimination_P2L} and \autoref{line:elimination_P2P}) in \autoref{algo:solve}, where $\tilde{p}$ denotes the parent of $p$. Second, $T_i$ updates operators associated with two neighbors of $i$ including $\MtoL_{pq}$, $\MtoL_{qp}$,  $\PtoL_{pq}$, $\MtoP_{qp}$, $\PtoP_{pq}$, and $\PtoP_{qp}$ in \autoref{algo:fact}. 

Overall, the neighbor lists of $i$ and $i'$ should not overlap, i.e., $\NL_i\cap\NL_{i'}=\emptyset$, which is guaranteed if
\begin{eqnarray*}
  \separation \ge 3.
  \label{eq:sep3}
\end{eqnarray*}

\begin{remark}{}
Proof of $\NL_i\cap\NL_{i'}=\emptyset$ when $\separation \ge 3$: since $3 \le \separation \le \dist(i, i') \le \dist(i, q) + \dist(q, q') + \dist(q', i')$, where $q \in \NL_i$ and $q' \in \NL_{i'}$, we have $\dist(q, q') \ge 1$, i.e., $\NL_i\cap\NL_{i'}=\emptyset$.
\end{remark}

\begin{remark}{}
\autoref{fig:sep3b} and \ref{fig:sep3} show two examples corresponding to the above two scenarios, and illustrate that $\NL_i\cap\NL_{i'}\not=\emptyset$ when $\separation=2$, while $\NL_i\cap\NL_{i'}=\emptyset$ when $\separation=3$.
\end{remark}

\begin{figure}[htb]
  \centering
  \begin{tabular}{cc}
  \includegraphics[height=.15\textheight]{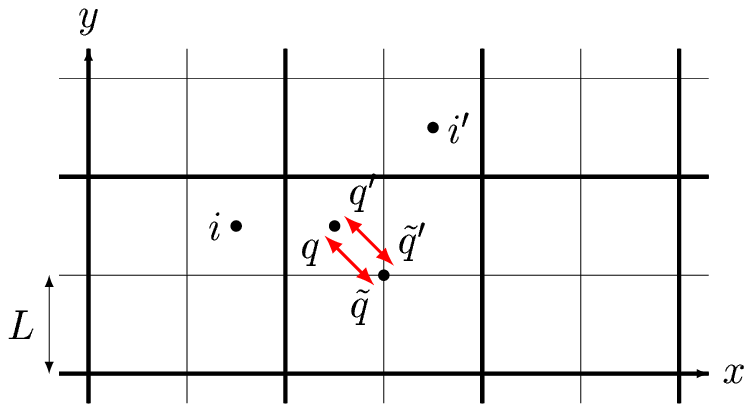}
    &\includegraphics[height=.15\textheight]{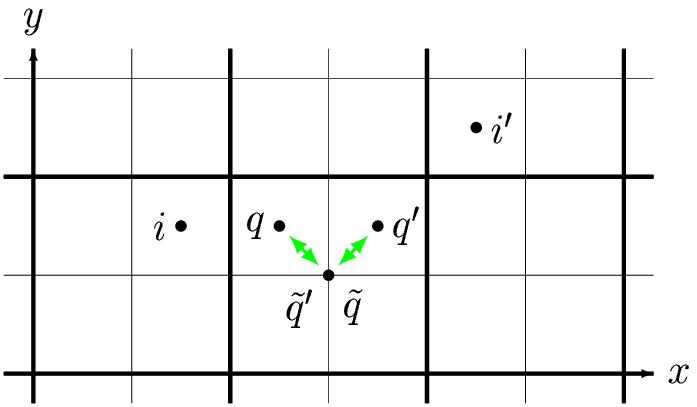}
  \end{tabular}
\caption{$T_i$ (resp. $T_{i'}$) updates operators (arrows) associated with $q$ and $\tilde{q}$ (resp. $q'$ and $\tilde{q}'$), where $\tilde{q}$ (resp. $\tilde{q}'$) is the parent, which corresponds to the center of the coarser box. (Left) When $\separation=2$, $q = q'$ and $\tilde{q} = \tilde{q}'$, so there is a race condition when $T_i$ and $T_{i'}$ update the corresponding operators (red arrows) in parallel. (Right) When $\separation=3$,  $\tilde{q} = \tilde{q}'$, but $q \not= q'$, so the corresponding operators (green arrows) can be updated in parallel.}
  \label{fig:sep3b}
\end{figure}


\begin{figure}[htb]
  \centering
  \begin{tabular}{cc}
  \includegraphics[height=.15\textheight]{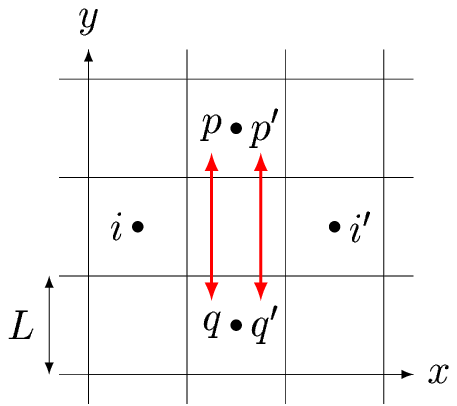}
    &\includegraphics[height=.15\textheight]{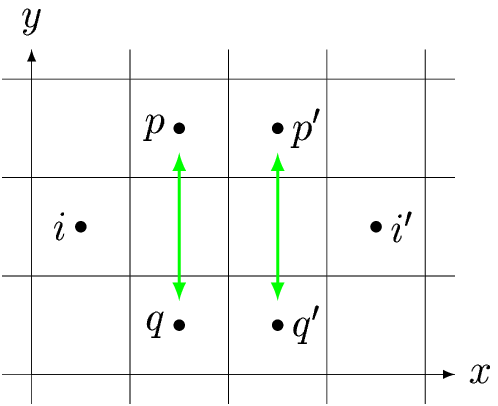}
  \end{tabular}
\caption{$T_i$ (resp. $T_{i'}$) updates operators (arrows) associated with $p$ and $q$ (resp. $p'$ and $q'$). (Left) When $\separation=2$, $p = p'$ and $q = q'$, so there is a race condition when $T_i$ and $T_{i'}$ update the corresponding operators (red arrows) in parallel. (Right) When $\separation=3$,  $p \not = p'$ and $q \not= q'$, so the corresponding operators (green arrows) can be updated in parallel.}
  \label{fig:sep3}
\end{figure}




\item \textbf{Neighbors' interaction list operations} 


This operation exists only in \autoref{algo:fact}, not in \autoref{algo:solve}. $T_i$ (resp. $T_{i'}$) updates $i$'s (resp. $i'$'s) neighbors' interaction list (underlined lines in \autoref{algo:fact}). Suppose $q \in \NL_i$ and $q' \in \NL_{i'}$. The condition that $q \not \in \IL_{q'} (\Leftrightarrow q' \not \in \IL_{q} )$ needs to be satisfied such that there is no race condition when $T_i$ updates $\IL_q$ and $T_{i'}$ updates $\IL_{q'}$ in parallel. In other words, the parents of $q$ and $q'$ cannot be neighbors. This is guaranteed if
  \begin{eqnarray*}
    \separation\ge 6.
    \label{eq:sep6}
  \end{eqnarray*}

\begin{remark}{}
Proof of $q \not \in \IL_{q'}$ when $\separation \ge 6$: since $6 \le \separation \le \dist(i, i') \le \dist(i, q) + \dist(q, q') + \dist(q', i')$, where $q \in \NL_i$ and $q' \in \NL_{i'}$, we have $\dist(q, q') \ge 4$, i.e., $q \not \in \IL_{q'}$ according to \autoref{eq:dist}.
\end{remark}

\begin{remark}{}
\autoref{fig:sep6} illustrates examples that $q \in \IL_{q'}$ when $\separation=5$, while  $q \not \in \IL_{q'}$ when $\separation=6$.
\end{remark}

\begin{figure}[htb]
  \centering
  \begin{tabular}{cc}
  \includegraphics[height=.175\textheight]{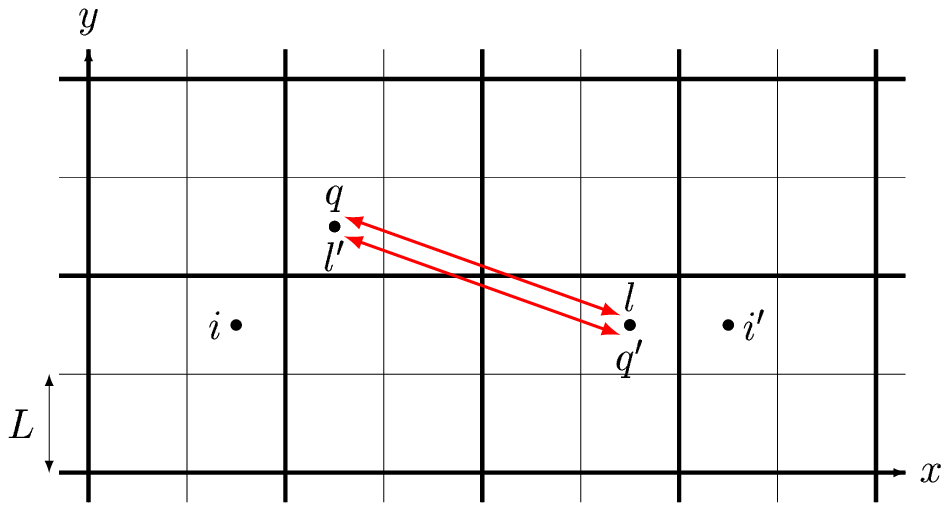}
    &\includegraphics[height=.175\textheight]{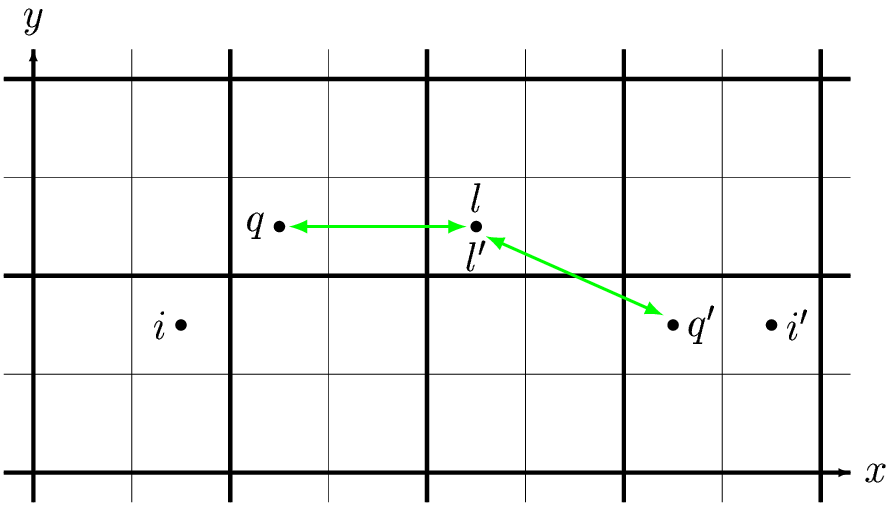}
  \end{tabular}
\caption{$T_i$ (resp. $T_{i'}$) updates $M2L_{ql}$ (resp. $M2L_{q'l'}$) where $q \in \NL_i, l \in \IL_q$ (resp. $q' \in \NL_{i'}, l' \in \IL_{q'}$). (Left) When $\separation=5$, $q = l'$, and $q' = l$, so there is a race condition when $T_i$ and $T_{i'}$ update $M2L_{ql}$ and $M2L_{q'l'}$ (red arrows), respectively, in parallel. (Right) When $\separation=6$, $q \not = l'$, and $q' \not= l$, so $M2L_{ql}$ and $M2L_{q'l'}$ (green arrows) can be updated in parallel.}
  \label{fig:sep6}
\end{figure}
\end{enumerate}

\subsection{Basic parallel algorithm ($\separation=6$)}\label{s:full}

As the above analysis shows, if nodes $i$ and $i'$ have the same color and $\separation=6$, two threads $T_i$ and $T_{i'}$ have no interference with each other in processing $i$ and $i'$, respectively. Therefore, we can replace the loop over nodes at the same level (line \autoref{line:fact_loop_i} in \autoref{algo:fact}, and lines \ref{line:elimination_loop_i} and \ref{line:substitution_loop_i} in \autoref{algo:solve}) with a double-loop over colors (in serial) and nodes that have the same color (in parallel). 

Specifically, we color the tree nodes at every level subject to the constraint that $\separation=6$ in \autoref{eq:sep} (\autoref{algo:group} in \autoref{s:grouping}). Suppose at level $\kappa$, there are $g_\kappa$ colors, and the tree nodes fall into $G_0, G_1, \ldots, G_{g_\kappa-1}$ groups according to their colors. Our basic parallel algorithm can be obtained by replacing
\begin{center}
\hspace*{20pt} \textbf{for} Node $i=0$ \textbf{to} $n_\kappa-1$
\end{center}
at line \autoref{line:fact_loop_i} in \autoref{algo:fact} and line \ref{line:elimination_loop_i} in \autoref{algo:solve} with
\begin{center}
\hspace*{20pt} \textbf{for} Color $I=0$ \textbf{to} $g_\kappa - 1$ \\
\hspace*{20pt} \quad \quad \quad \, \#pragma omp parallel for\\
\hspace*{20pt} \quad \textbf{for all} Node $i \in G_I$
\end{center}
and replacing
\begin{center}
\hspace*{20pt} \textbf{for} Node $i=n_\kappa-1$ \textbf{to} $0$
\end{center}
at line \ref{line:substitution_loop_i} in \autoref{algo:solve} with
\begin{center}
\hspace*{20pt} \textbf{for} Color $I=g_\kappa - 1$ \textbf{to} $0$ \\
\hspace*{20pt} \quad \quad \quad \, \#pragma omp parallel for\\
\hspace*{20pt} \quad \textbf{for all} Node $i \in G_I$
\end{center}
using the ``parallel for'' pragma in OpenMP.


Note that we have to use the same coloring ($\separation$) in \autoref{algo:fact} and \ref{algo:solve} because the ordering in the solve needs to be consistent with that in the factorization; in particular for solving dense linear systems, the ordering of the solve is exactly the reverse of the factorization. Therefore, we cannot apply $\separation=3$ for \autoref{algo:solve} despite the fact thatthe solve is irrelevant to the data dependency of (c).




\subsection{Improved parallel algorithm ($\separation=3$)}\label{s:part}

To improve the efficiency of the basic parallel algorithm, we consider using a smaller $\separation$ in the coloring algorithm at the cost of serializing a part of the algorithm. However, $\separation \le 2$ leads to race conditions in updating neighbors' operators and neighbors' interaction list operators, meaning the factorization and solve are serialized entirely. So, let us consider $\separation=3$, in which case $T_i$ and $T_{i'}$ can work independently in \autoref{algo:solve}, and they interfere with each other only for neighbors' interaction list operations in \autoref{algo:fact}. Therefore, our improved parallel algorithm is based on the coloring with constraint $\separation=3$, and \textit{serializes} the neighbors' interaction list operations (underlined in \autoref{algo:fact}).

Specifically, we still color the tree nodes at every level (\autoref{algo:group} in \autoref{s:grouping}), but with the constraint that $\separation=3$ in \autoref{eq:sep} as apposed to the $\separation=6$ constraint used in the basic parallel algorithm. With the coloring results, we follow the previous parallel algorithm and replace the serial loop over nodes at every level in \autoref{algo:fact} and \ref{algo:solve} with a double-loop, where the outer loop is over all colors in serial and the inner loop is over nodes of the same color in parallel. In addition, we serialize updating neighbors' interaction list operators (underlined in \autoref{algo:fact}) to avoid race condition. This serialization can be implemented with OpenMP locks~\cite{OpenMP} or with OpenMP's critical constructor ``\#pragma omp critical'', which allows only one thread in the critical region at a time. Our implementation adopts the latter for the underlined statements in \autoref{algo:fact}.

Compared with the coloring results ($\separation=6$) used in the basic parallel algorithm, the improved parallel algorithm would have less number of colors and larger group sizes on average for every color, which implies less synchronization cost and more concurrency. The extra serialization cost in the improved parallel algorithm is expected to be small because empirically updating neighbors' interaction list is a small fraction of the total runtime.

\subsection{Coloring algorithm}\label{s:grouping}

Given the tree nodes at the level $\kappa$ in the FMM-tree, we want to assign colors to them so that \autoref{eq:sep} holds for a prescribed $\separation$. The number of colors used is denoted by $g_\kappa$, and the nodes can be classified into groups $G_0, G_1, \ldots, G_{g_\kappa-1}$ based on their colors.

\begin{figure}[!htbp]
  \centering
  \begin{subfigure}{0.5\textwidth}
	\centering
  	\includegraphics[width=.65\textwidth]{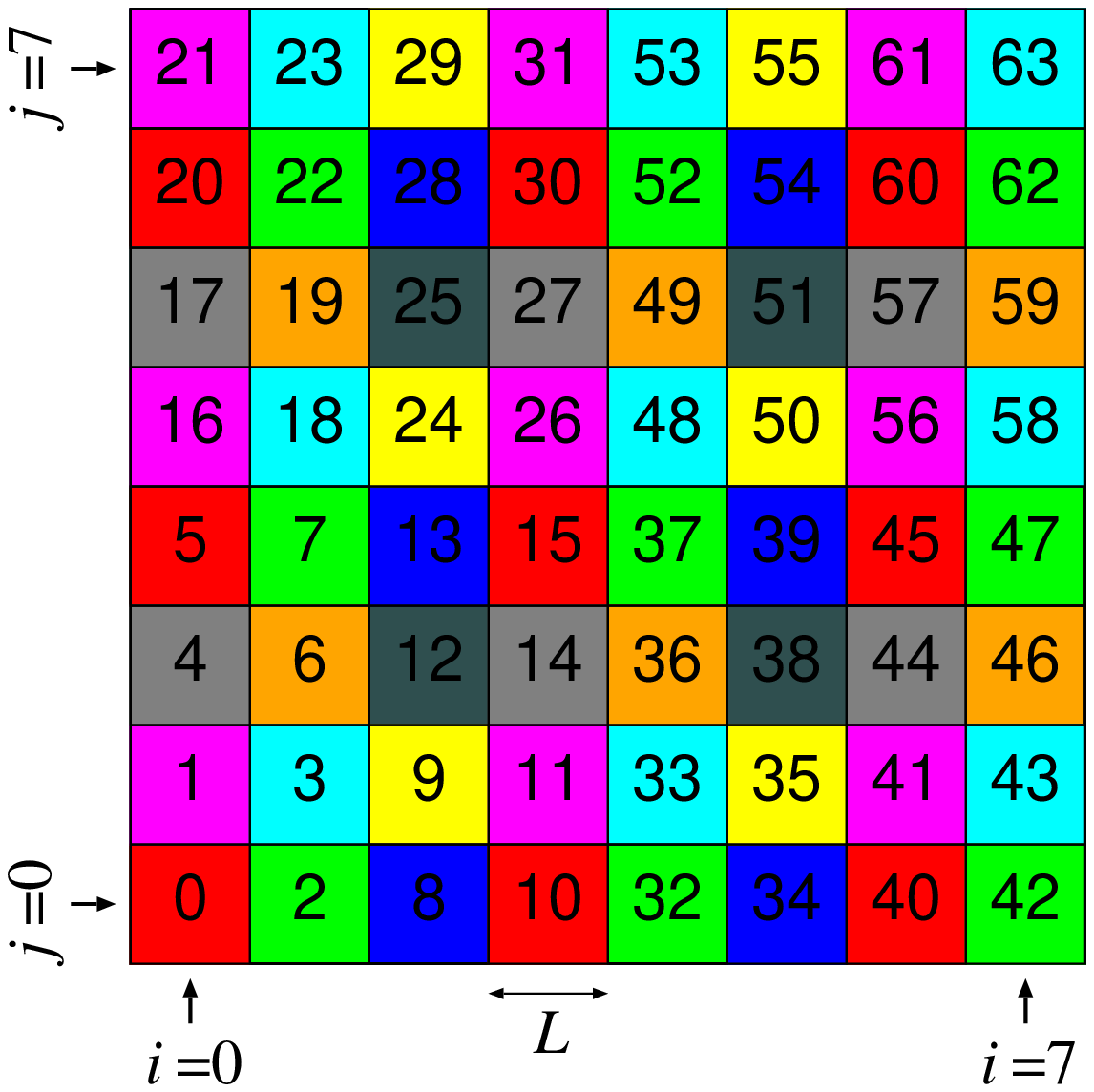}
	\caption{All nodes are colored}
  \end{subfigure}%
  \begin{subfigure}{0.5\textwidth}
	\centering
  	\includegraphics[width=.66\textwidth]{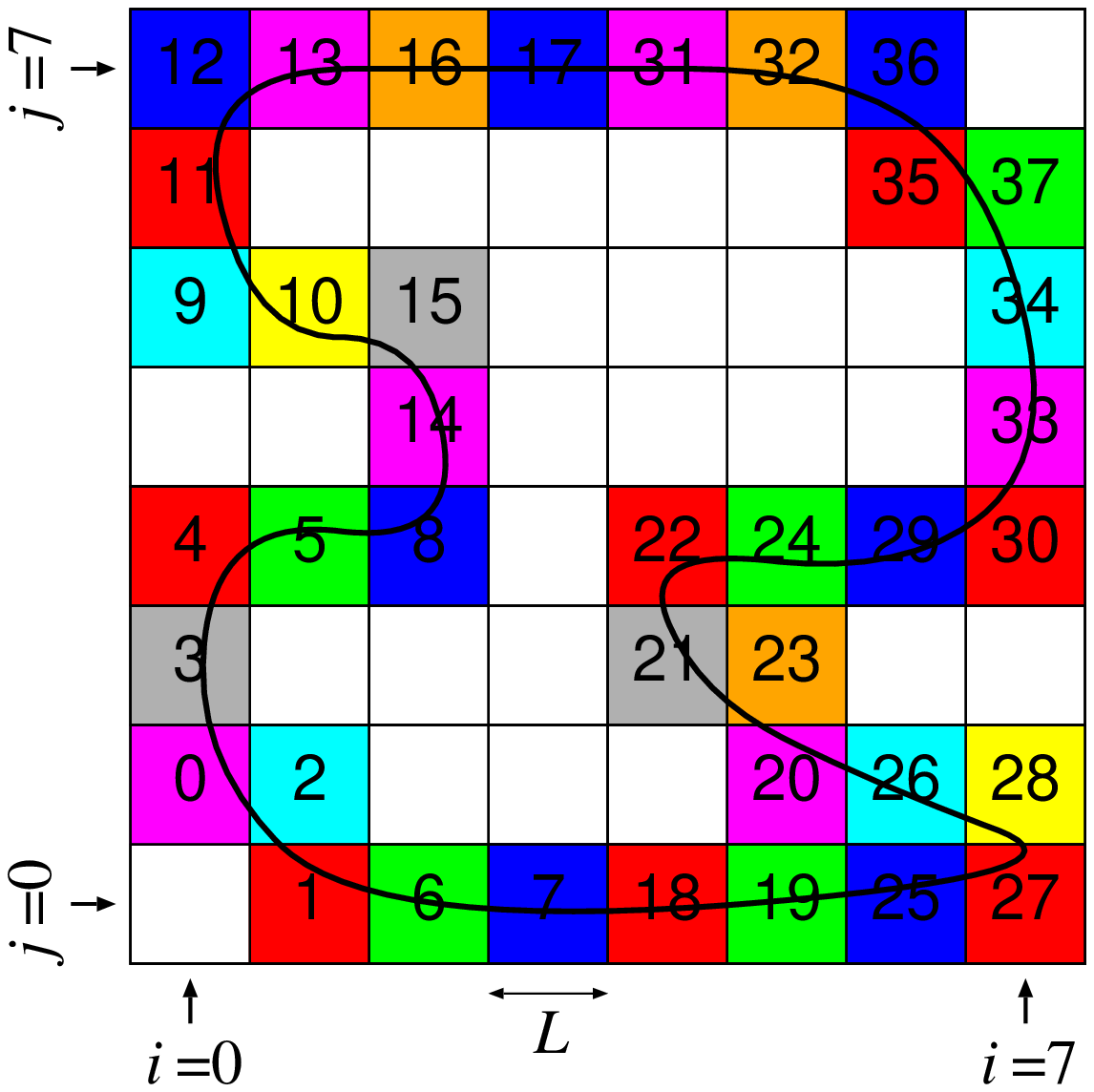}
  	\caption{Boundary nodes are colored}
  \end{subfigure}
  \caption{Coloring results of \autoref{algo:group} in 2D for tree nodes at level $\kappa=3$ and $\separation=3$. (a) The optimal coloring is achieved with $\separation^2=9$ colors. Every color corresponds to a sub-grid of stride 3 in both directions. (b) Only nodes corresponding to the domain boundary are colored. $g_\kappa = 8$ colors are used.}
  \label{fig:group_example}
\end{figure}

When the FMM-tree is complete, the nodes at the same level lie on a regular grid. It is easy to see that the optimal coloring has $\separation^3$ colors, and nodes of the same color lie on a sub-grid of stride $\separation$, as shown in \autoref{fig:group_example} (a). In general, some of the nodes may be empty and do not need to be colored. In the context of the BEM, only tree nodes corresponding to boundary elements need to be colored. Such an example is shown in \autoref{fig:group_example} (b). 

The idea of our greedy algorithm is as follows. We find the first node that is not empty nor colored, and we color it. Then, we try to find other nodes that can be assigned the same color. To find the next node, we first make a stride of $\separation$ and search for a non-empty and un-colored node. Once we find the second node, we make a stride of $\separation$ again and continue the search. After we have exhausted all the remaining nodes, we start over for a new color. This algorithm is shown in \autoref{algo:group}.


We note that the computational cost of \autoref{algo:group} is $O(n_\kappa)$. Since up to $\separation^3$ groups are created at line \ref{line:new}, the outer triple-loop over $k_0$, $j_0$, and $i_0$ is executed up to $\separation^3$ times. Since each length of the loops over $k$, $j$, and $i$ is up to $2^\kappa/\separation$, the computational complexity is effectively $\separation^3\times(2^\kappa/\separation)^3=n_\kappa$. In fact, we observed the time to group $n_\kappa$ nodes is proportional to $n_\kappa$ in the two examples in~\autoref{s:sphere} and \autoref{s:office}, where $n_\kappa \lesssim 116$k. In addition, the elapsed time was less than 0.1 seconds in the case of the maximum $n_\kappa$ for each example.

We also note that \autoref{algo:group} does not guarantee to minimize the number of colors or maximize the (average) number of nodes per group. This is due to the fixed stride of $\separation$ (lines \ref{line:g_stride_i}, \ref{line:g_stride_j}, and \ref{line:g_stride_k}). For example, if we look back all the nodes already colored, we would be able to find more nodes in a group but lead to more computational cost.

\def\MYIF{\textbf{if}}
\def\MYTHEN{\textbf{then}}
\def\MYELSE{\textbf{else}}
\def\MYENDIF{\textbf{end if}}
\def\MYAND{\textbf{\&}}


\begin{algorithm}[htbp]
  \caption{Coloring algorithm. Input: all tree nodes at level $\kappa$; output: color of every node and the number of colors $g_\kappa$. Notation: $m=2^\kappa$ is the number of nodes at each dimension. Based on colors, nodes can be classified into $g_\kappa$ groups $G_0, G_1, \ldots, G_{g_\kappa-1}$.}

  \label{algo:group}
  \begin{algorithmic}[1]
    \STATE Initialize the first color $g_\kappa=0$.
    \FOR{$k_0 = 0$ \TO $m-1$}
    \FOR{$j_0 = 0$ \TO $m-1$}
    \FOR{$i_0 = 0$ \TO $m-1$}
    \STATE Denote the $(i_0,j_0,k_0)$-th node as $u$.
    \IF{$u$ is not empty \MYAND\ $u$ is not colored} \label{line:if}
    \STATE Assign color $g_\kappa$ to node $u$.  \blue{// $u$ is the first node with color $g_\kappa$} \label{line:new}
    \STATE $j\_stride=true$, $k\_stride=true$ \blue{// next node is at least one stride away}

    \STATE $k = k_0$
    
    \STATE  \blue{// check if node $(i,j,k)$ can be colored $g_\kappa$} 
     
    \WHILE{$k < m$}
    
    \STATE \MYIF\ $k==k_0$\ \MYTHEN\ $j=j_0$\ \MYELSE $j=0$\ \MYENDIF

    \WHILE{$j < m$}
    
    \STATE \MYIF\ $j==j_0$ \MYAND\ $k==k_0$\ \MYTHEN\ $i = i_0 + \separation$\ \MYELSE\ $i = 0$ \MYENDIF
    
    \WHILE{$i < m$}
    \STATE Denote the $(i,j,k)$-th node as $v$.
    \IF{$v$ is not empty \MYAND\ $v$ is not colored}
    \STATE Assign color $g_\kappa$ to node $v$. 
    \STATE \blue{// next node is at least one stride away}
    \STATE $i = i + \separation$, $j\_stride=true$, $k\_stride=true$ \label{line:g_stride_i}
    \ELSE
    \STATE $i =i+1$
    \ENDIF \blue{\; // $v$}
    \ENDWHILE  \blue{\; // $i$}
    
    \STATE \MYIF\ $j\_stride==true$\ \MYTHEN\ $j=j+\separation$ \ \MYELSE\ $j=j+1$ \MYENDIF \label{line:g_stride_j}
    \STATE $j\_stride = false$

    \ENDWHILE \blue{\; // $j$}
    
    \STATE \MYIF\ $k\_stride==true$\ \MYTHEN\ $k=k+\separation$\ \MYELSE\ $k=k+1$ \MYENDIF \label{line:g_stride_k}
    \STATE $k\_stride = false$

    \ENDWHILE \blue{\; // $k$}

    \STATE $g_\kappa =g_\kappa+1$  \label{line:g}

    \ENDIF  \blue{\; // $u$}
    \ENDFOR \blue{\; // $i_0$}
    \ENDFOR \blue{\; // $j_0$}
    \ENDFOR \blue{\; // $k_0$}
  \end{algorithmic}
\end{algorithm}

\section{Application of IFMM to FMBEM}\label{s:preconditioner}

We apply the proposed parallel IFMM to an iterative BEM as the preconditioner. In this section, we first overview the algorithm of the BEM. Then, we define two types of the IFMM-based preconditioners, i.e., mIFMM and nIFMM. Last, we explain the standard preconditioner, that is, the leaf-based block-diagonal preconditioner and its enhancement to be compared with the IFMM-based preconditioners.

\subsection{FMBEM}\label{s:bem}

Following the previous study~\cite{takahashi2017}, we deal with the BEM for 3D Helmholtz equation. Specifically, we solve the acoustic scattering problems with the combined (or Burton-Muller type) boundary integral equations (CBIE)~\cite{1971_Burton_Miller}. We use the preconditioned GMRES~\cite{2003_Saad_book} to iteratively solve the discretized CBIE or the resulting linear system
\begin{eqnarray}
  Ax=b,
  \label{eq:axb_bem}
\end{eqnarray}
where $A\in\bbbc^{N\times N}$ and $x,b\in\bbbc^N$. Here, $N$ stands for the number of piece-wise constant elements to discretize the surface of a given model (scatterer). To accelerate the matrix-vector product for $A$, we use the low-frequency FMM (LFFMM)~\cite{1995_Epton,2002_Nishimura,2009_Liu_book,2011_Liu,2016_Takahashi,darv04b}.  The overall algorithm of our FMBEM is described in \autoref{algo:fmbem}. 

\begin{algorithm}[h]
  \caption{FMBEM using the preconditioned GMRES.}
  \label{algo:fmbem}
  \begin{algorithmic}[1]
    \STATE Input a boundary element mesh, boundary conditions, parameters, etc.
    \STATE Build an FMM's adaptive octree with a depth $\lFMM$; specifically, see \autoref{s:num_setting}.
    \STATE Precompute all the FMM operators, viz., P2M, M2M, M2L, L2L, L2P, and P2P. \label{line:precomp}
    \STATE Compute the RHS vector $b$ in \autoref{eq:axb_bem} by the matrix-vector product routine.
    \IF{mIFMM or nIFMM is used}\label{line:init_start}
    \STATE Perform the hierarchical domain decomposition (\autoref{s:hdd}), which builds a uniform octree with a depth $\ell$.
    \STATE Initialize the tree data structure (\autoref{s:init_tree_data}), which constructs the extended sparse matrix $\bar{A}$ of $A$ in \autoref{eq:axb_bem} with partially utilizing the FMM operators stored at line \ref{line:precomp}. \label{line:reuse} \label{line:fmbem_init}
    \STATE Factorize $\bar{A}$ according to \autoref{algo:fact}. \label{line:fmbem_fact}
    \ELSIF{BD is used}
    \STATE Precompute the LU factorization of the self P2P operators, i.e., $\PtoP_{ii}$, for every leaf node $i$ (in terms of the adaptive octree). This computation is fully parallelized with respect to $i$.
    \ENDIF\label{line:init_end}
    \STATE Perform GMRES to solve \autoref{eq:axb_bem} with the initial guess of $x=0$.
    \STATE Output the solution $x$.
  \end{algorithmic}
\end{algorithm}


The preconditioning is performed through \autoref{algo:precon}. Specifically, the algorithm computes $Mr$, where $M$ denotes a preconditioner such as $M\approx A^{-1}$ and $r$ is a known vector at a given iteration step; therefore, computing $Mr$ means to solve $Av=r$ approximately for $v$. As $M$, we will use nIFMM, mIFMM, and BD introduced below. 

\begin{algorithm}[h]
  \caption{Preconditioning routine to solve $M^{-1}v=r$ for $v$ when a vector $r$ is given.}
  \label{algo:precon}
  \begin{algorithmic}[1]
    \IF{mIFMM or nIFMM is used}
    \STATE Solve the extended sparse system according to \autoref{algo:solve}, where the input vector $b$ and the output vector $x$ are regarded as $r$ and $v$, respectively.
    \ELSIF{BD is used}
    \STATE Solve $v_i$ from $\PtoP_{ii}v_i=r_i$ with the precomputed LU factorization of $\PtoP_{ii}$ for every leaf node $i$ at level $\lBD$, where $v_i$ and $r_i$ denote the part of $v$ and $r$ regarding $i$, respectively. This computation is fully parallelized with respect to $i$.
    \ENDIF
    \RETURN $v$.
  \end{algorithmic}
\end{algorithm}

\subsection{IFMM-based preconditioners: nIFMM and mIFMM}\label{s:precon_ifmm}


In the preconditioning routine, we refer to the preconditioner $M$ as \textbf{IFMM-based preconditioner} if $v$ is obtained as a part of $\bar{v}$ that is approximately solved from the IFMM-extended sparse system $\bar{A}\bar{v}=\bar{r}$, where $\bar{r}$ is the extended vector of a given $r$. To use this preconditioner, we first construct $\bar{A}$ and then factorize it according \autoref{algo:fact} or its parallel version before starting GMRES (see lines~\ref{line:fmbem_init} and \ref{line:fmbem_fact} in \autoref{algo:fmbem}). Subsequently, \autoref{algo:solve} or its parallel version is performed when the preconditioning routine is called from GMRES (see \autoref{algo:precon}).

We will denote an IFMM-based preconditioner as \textbf{nIFMM} if IFMM is parallelized with respect to ``n''odes according to \autoref{s:full} (where $\separation=6$ is employed) or \autoref{s:part} (where $\separation=3$). In addition, we will denote another IFMM-based preconditioner as \textbf{mIFMM} if IFMM is parallelized in the grain of individual ``m''atrix operations by means of a multi-threaded linear algebra package; we actually use Intel MKL~\cite{Intel_MKL} through Eigen~\cite{eigenweb}.

We construct the IFMM with the LFFMM; in general, we may handle the IFMM with a different FMM (e.g., the black-box FMM~\cite{fong2007}) from the FMM for the matrix-vector product, i.e., LFFMM in this study. Although it would be simple to manipulate both LFFMMs with the same hierarchy (of boundary elements), the hierarchy of the LFFMM for the matrix-vector products employs the adaptive octree differently from the uniform octree of the IFMM (recall \autoref{s:hdd}). Nevertheless, to share the information as much as possible, we decided that both root nodes have the same position and size. Then, the FMM operators can be shared by the nodes of both FMMs especially in upper levels (see line~\ref{line:reuse} in \autoref{algo:fmbem}). This (irregular) combination of hierarchies was actually used in the previous work~\cite{takahashi2017}.

Yet, the depth of the IFMM's hierarchy (i.e., $\ell$) is not necessarily the same as that of the FMM's hierarchy (i.e., $\ell_{\rm FMM}$), although $\ell\equiv\ell_{\rm FMM}$ was supposed in \cite{takahashi2017}. We will determine the value of $\ell_{\rm FMM}$ by a conventional manner, as described in \autoref{s:num_setting}. Meanwhile, we regard $\ell$ as a tunable parameter, which can affect the performance of nIFMM and mIFMM.

Basically, if we let $\ell$ be less than $\lFMM$, we no longer perform the factorization at the omitted levels $\ell+1$ to $\lFMM$. In addition, we can handle a smaller number of nodes at level $\ell$ than $\lFMM$. The downside is that the matrix manipulation (e.g., low-rank compression) can require more computational cost because the size of individual matrices (FMM operators) increases as $\ell$ decreases; this is because the size is determined by the formula in \autoref{eq:p} in \autoref{s:num_setting}. Further, the initialization cost can become large (respectively, small) if the IFMM cannot (respectively, can) fully reuse the FMM operators, especially the P2P operators, precomputed at leaf nodes of the FMM's hierarchy (see line~\ref{line:reuse} in \autoref{algo:fmbem}, again).

Another parameter of nIFMM is the top level $\tau$ ($\in[2,\ell]$). Namely, the node-based parallelism is applied to level $\tau$ to $\ell$, while the matrix-based parallelism is applied to level 2 to $\tau-1$. By switching the parallelism from the node-based to the matrix-based, we can expect a higher performance at coarse (upper) levels. This is because individual matrix is larger at coarse levels than fine levels and, thus, can be processed more efficiently with multiple threads.

For conciseness, we denote the nIFMM using the parameters $\separation$ (``sep''aration), $\tau$ (``top'' level), and $\ell$ (``bot''tom level) by \textbf{nIFMMsep$\separation$top$\tau$bot$\ell$} or \textbf{nIFMMs$\separation$t$\tau$b$\ell$} for short. Regarding mIFMM, we use the term \textbf{mIFMMbot$\ell$} or \textbf{mIFMMb$\ell$} to specify the bottom level $\ell$. \autoref{tab:names} summarizes the naming scheme of the IFMM-based preconditioners as well as the (extended) block-diagonal preconditioner, which will be explained in the next section.

\subsection{Block-diagonal preconditioner: BD}\label{s:precon_bd}

We will compare mIFMM and nIFMM with the leaf-based block-diagonal preconditioner (referred to as \textbf{BD}), which is frequently used for FMBEM~\cite{2009_Liu_book}. The corresponding preconditioning matrix $M$ is computed as a block-diagonal matrix consisting of the inverse of the self P2P operators (i.e., $\PtoP_{ii}$) in terms of leaf nodes (see \autoref{algo:precon}), where the FMM that BD relies on is nothing but the LFFMM for the matrix-vector product. Each self P2P operator is LU-decomposed at the initialization stage (see \autoref{algo:fmbem}). It should be noted that BD is fully parallelizable with respect to leaf nodes in each level in both initialization and application (preconditioning) stages.

We can improve the performance of BD by selecting its bottom level (depth), denoted by $\lBD$, where we actually compute the diagonal blocks of $M$. If we let $\lBD<\lFMM$, the corresponding $M$ can have a wider bandwidth and become a better approximation of $A^{-1}$. Then, it can reduce the number of iterations until convergence. On the other hand, the cost and memory for factorizing the individual diagonal blocks becomes more expensive at the initial stage. We denote the enhanced BD using the bottom level of $\lBD$ by \textbf{BDbot$\lBD$} or \textbf{BDb$\lBD$}.

\begin{table}[h]
  \centering
  \caption{List of the preconditioners. The value of $\lFMM$ is determined as in \autoref{s:num_setting}.}
  \label{tab:names}
  \begin{tabular}{|c|c|p{100pt}|}
    \toprule
    Name & Name specifying parameters & \\
    \midrule
    nIFMM  & \begin{tabular}{c} nIFMMsep$\separation$top$\tau$bot$\ell$\quad or\quad nIFMMs$\separation$t$\tau$b$\ell$ \\ (where $\separation=3$ or $6$ and $2\le\tau\le\ell\le\lFMM$) \end{tabular} & Applies the matrix-based parallelism to level 2 to $\tau-1$ and the node-based parallelism to level $\tau$ to $\ell$.\\
    \midrule
    mIFMM  & \begin{tabular}{c} mIFMMbot$\ell$\quad or\quad mIFMMb$\ell$ \\ (where $2\le\ell\le\lFMM$) \end{tabular} & Applies the matrix-based parallelism to all the levels, i.e. 2 to $\ell$.\\
    \midrule
    BD & \begin{tabular}{c} BDbot$\lBD$\quad or\quad BDb$\lBD$ \\ (where $2\le\lBD\le\lFMM$) \end{tabular} & Handles nodes at level 2 to $\lBD$.\\
    \bottomrule
  \end{tabular}
\end{table}

\section{Numerical experiments}\label{s:num}

In comparison with mIFMM and BD, we will assess the proposed parallel IFMM-based preconditioner, i.e., nIFMM, through solving two acoustic scattering problems.

\subsection{Details of GMRES, FMM,  IFMM, and computers}\label{s:num_setting}

We used the restart GMRES, i.e., GMRES($m$)~\cite{2003_Saad_book}, but no restart took place in the following tests since we used a sufficiently large restart parameter $m$, i.e., $m=1000$ for the IFMM-based preconditioners and $m=3000$ for BD. The iteration was terminated when the relative residual $\norm{Ax-b}_2/\norm{b}_2$ became smaller than the predefined number ``$tol$''. We set $10^{-5}$ to $tol$. We note that our GMRES program is not parallelized except for the preconditioning (\autoref{algo:precon}) and matrix-vector routines, which is performed by the parallelized LFFMM (see below). The parallelization can lead to the reduction of the total computation time if the number of iterations is large.

Regarding the (LF)FMM, its performance can be controlled by two parameters. The first one is the maximum number of boundary elements per leaf node, denoted by $\mu$. Once $\mu$ is given, the depth of FMM's octree is determined as $\ell_{\rm FMM}$ ($\ge 2$). The second parameter is the precision factor $\delta$ to determine the number $p$ of the multipole and local coefficients through the empirical formula~\cite{1993_Coifman}
\begin{eqnarray}
  p:=\sqrt{3}Lk+\delta\log\left(\sqrt{3}Lk+\pi\right),
  \label{eq:p}
\end{eqnarray}
where $k$ is a prescribed wavenumber. We set $7$ to $\delta$ in all the computations. We note that the row and/or column size of an initial (uncompressed) FMM operator is typically $p^2$ (not $p$). Also, we note that our LFFMM code is parallelized with respect to the loop over nodes in each level.

The major parameter of the IFMM is the relative error $\varepsilon$ that determines the accuracy of the low-rank compression, for which we used the randomized SVD~\cite{2011_Halko}. In general, a smaller $\varepsilon$ makes the elapsed times of factorization and solve phases longer but can reduce the number of iterations~\cite{takahashi2017}. We mainly used $\varepsilon=10^{-3}$ and considered $10^{-1}$ and $10^{-2}$ for comparison.

In the case of the first example in \autoref{s:sphere}, we used 16 cores in a desktop PC (which consists of two Intel Xeon Gold 6134 CPUs) and \unit{512}{GB} memory in total. As a compiler, gcc-7.3.1 was used with the optimizing flags ``-O3 -march=native -mtune=native''\footnote{In addition, we built the IFMM program with the Eigen's macro ``EIGEN\_USE\_MKL\_ALL'' to parallelize each matrix operation by Intel MKL~\cite{Intel_MKL} outside parallel regions. This actually improved the performance of the related codes.}. In the second example in \autoref{s:office}, we needed a much memory because of larger $N$ and $k$ than the first case. Therefore, we used a computing node with \unit{3}{TB} memory and 56 cores (in four Intel Xeon E5-4650v4 CPUs). The compiler was gcc-7.3.0 with the aforementioned flags\footnote{Since both computers are based on the NUMA architecture, we considered to bind the OpenMP threads to specific computing cores. We simply bound the $i$-th thread to the $i$-th core. This improved the timing result in practice.}.

\subsection{Test problem --- sphere model}\label{s:sphere}

\subsubsection{Problem statement}

We considered an external Neumann problem for an acoustically hard sphere (with the radius of 0.5 and the center at the origin) irradiated by a point source $f(\fat{x})$ at $\fat{s}=(0.0,0.0,0.8)^{\mathrm{T}}$, i.e., $f(\fat{x}):=\exp(\mathrm{i} k\abs{\fat{x}-\fat{s}})/(k\abs{\fat{x}-\fat{s}})$. We discretized the surface of the sphere with 1003520 triangular piece-wise constant boundary elements. The exact solution is available from the reference~\cite[Section 10.3.1]{Bowman}. In what follows, we considered $k=4$, $8$, $16$, and $32$.; the diameter of the sphere is about five times larger than the wavelength when $k=32$.

\subsubsection{Survey of the best BD, mIFMM, and nIFMM}\label{s:sphereii_best}

First of all, we searched the best value of the parameter $\mu$ (recall \autoref{s:num_setting}) so that the total computation time of the FMBEM based on the standard BD was minimized. As a result, $\mu=40$ was nearly the best. Then, the corresponding depth $\lFMM$ was $7$.

Consequently, using $\mu=40$, we surveyed the best (fastest) BD, mIFMM, and nIFMM by varying the parameters $\lBD$, $\ell$, and $\tau$, where the separation parameter $\separation$ was fixed to $3$ (\autoref{s:part}). As a result, we determined \textbf{BDbot5}, \textbf{mIFMMbot6}, and \textbf{nIFMMsep3top3bot6} as the best ones in this example. The details are described in Section~1 of the supplementary material.

\autoref{tab:stat_sphereii10} shows the statistics of the hierarchies used in the following computations.  We can observe that the number of nodes of the FMM and IFMM (in the second and fourth columns, respectively) is $\sim 4^\kappa$ at level $\kappa$, which is because the distribution of boundary elements is planar in 3D. Regarding nIFMM, the statistics of the grouping with $\separation=3$ is shown in the fifth to eighth columns. We can see that the group size at level $\kappa$ is $\sim 4^\kappa$ in average. In addition, $p\sim\kappa^{-1}$ is observed in the ninth and tenth columns.

\begin{table}[h]
  \def\prave#1{{\nprounddigits{0}\numprint{#1}}}
  \centering
  \caption{Statistics of the \textbf{sphere model}. ``\#nodes'': number of nodes of the FMM's adaptive octree when $\mu=40$. ``\#ele/node'': average number of elements per node. ``IFMM's \#node'': number of nodes of the IFMM's uniform octree when $\mu=40$. ``\#groups'': number of nIFMM's groups when $\separation=3$. ``Min'', ``Ave'', and ``Max'': minimum, average, and maximum group size. ``$p^2$'': the squared number of multipole and local coefficients. Note that this statistics is independent of the wavenumber $k$ as well as the precision parameter $\varepsilon$.}
  \label{tab:stat_sphereii10}
  %
  \small
  \begin{tabular}{|c|r|r|r|r|r|r|r|r|r|}
    \toprule
    Level  & \multicolumn{2}{c|}{FMM} & IFMM & \multicolumn{4}{c|}{nIFMM} & \multicolumn{2}{c|}{$p^2$}\\
    $\kappa$ & \#nodes & \#ele/node & \#nodes & \#groups & Min & Ave & Max & $k=4$ & $k=32$\\
    \midrule
    2 & 56 & \prave{17920.000000} & 56  & 19 & 2 & \prave{2.947368} & 8 & 144 & 1089\\
    3 & 272 & \prave{3689.411765} & 272 & 26 & 7 & \prave{10.461538} & 16 & 100 & 529\\
    4 & 1160 & \prave{865.103448} & 1160 & 27 & 8 & \prave{42.962963} & 70 & 81 & 256\\
    5 & 4694 & \prave{213.563698} & 4742 & 27 & 84 & \prave{175.629630} & 245 & 64 & 144\\
    6 & 17640 & \prave{56.232766} & 18488 & 27 & 305 & \prave{684.740741} & 973 & 64 & 100\\
    7 & 58730 & \prave{15.263988} & 70896 & 27 & 1173 & \prave{2625.777778} & 3712 & 64 & 81\\
    \bottomrule
  \end{tabular}
\end{table}

\subsubsection{Comparison under the best setting}\label{s:sphereii_comp3}

\begin{figure}[h]
  \centering

  \iftrue 


  \begin{tikzpicture}
    \begin{axis}[
      xlabel={Wavenumber $k$},
      ylabel style={align=center},
      ylabel={Speedup of nIFMMs3t3b6\\relative to  BDb5 and mIFMMb6},
      xmode=log,
      xmin=1, xmax=100,
      ymin=0.0, ymax=3.0, ytick={0.0,0.5,...,3.0},
      legend pos=south east,
      legend cell align={left},
      ymajorgrids=true,
      grid style=dashed,
      legend entries={BDb5,mIFMMb6},
      ]
      
      \addplot table [x=k, y expr=\thisrow{nIFMM}/\thisrow{BD}] {./181031/sphereii/fig-tsp-nIFMM-sphereii10-delta7-epsilon1e-3-relto-BD.table};
      \addplot table [x=k, y expr=\thisrow{nIFMM}/\thisrow{mIFMM}] {./181031/sphereii/fig-tsp-nIFMM-sphereii10-delta7-epsilon1e-3-relto-mIFMM.table};
      
    \end{axis}
  \end{tikzpicture}

\else

  \includegraphics[width=.5\textwidth]{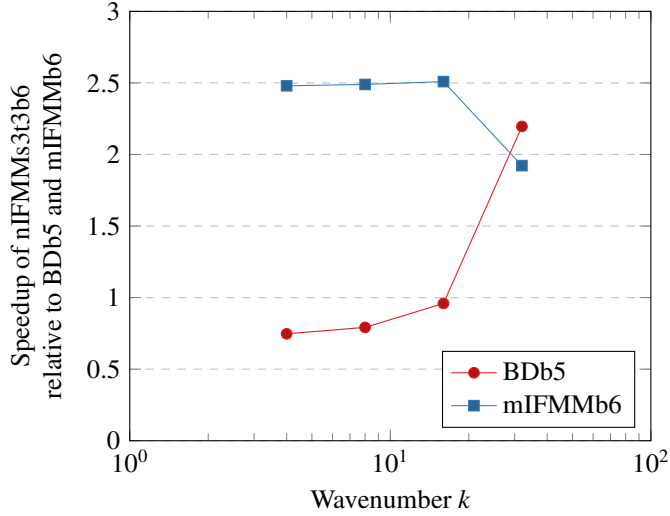}

\fi
  \caption{Speedup of the best nIFMM (i.e., nIFMMsep3top3bot6) relative to the best BD (i.e., BDbot5) and mIFMM (i.e., mIFMMbot6) in terms of the total computation time for the sphere model.}
  \label{fig:tsp-nIFMM}
\end{figure}

\autoref{fig:tsp-nIFMM} shows the speedup (in terms of the total computation time) of the best nIFMM with the best BD and mIFMM. Although the best nIFMM was slower than or comparable to the best BD for low wavenumbers, it achieved 2.2 times speedup at the largest $k$. It should be noted that the performance of BD can be improved by parallelizing the present GMRES program. Even if this is done successfully, the best nIFMM would remain faster than the best BD at $k=32$; this is indicated in Tables~I and III of the supplementary material.

In addition, the best nIFMM was about two times faster than the best mIFMM for any $k$. The drop at $k=32$ can be explained by the fact that the performance of mIFMM can increase with $k$, which will be examined in \autoref{s:sphereii_nIFMM_vs_mIFMM}.
 
\subsubsection{Comparison of node- and matrix-based parallelizations}\label{s:sphereii_nIFMM_vs_mIFMM}

\begin{figure}[h]
  \centering

  \iftrue 
  
  
  \begin{tikzpicture}
    \begin{axis}[
      xlabel={Wavenumber $k$},
      ylabel={Speedup},
      xmode=log,
      xmin=1, xmax=100,
      ymin=0.0, ymax=16.0, ytick={0.0,2.0,...,16.0},
      legend pos=south west,
      legend cell align={left},
      ymajorgrids=true,
      grid style=dashed,
      legend entries={factorization phase, solve phase}
      ]

      \addplot table [x=k, y=sp_ifmm_elim] {./181031/sphereii//fig-nIFMM_relto_mIFMM-sphereii10-delta7-epsilon1e-3.table}; 
      \addplot table [x=k, y=sp_msolve_per_iter] {./181031/sphereii//fig-nIFMM_relto_mIFMM-sphereii10-delta7-epsilon1e-3.table}; 
      
    \end{axis}
  \end{tikzpicture}

  \else

  \includegraphics[width=.5\textwidth]{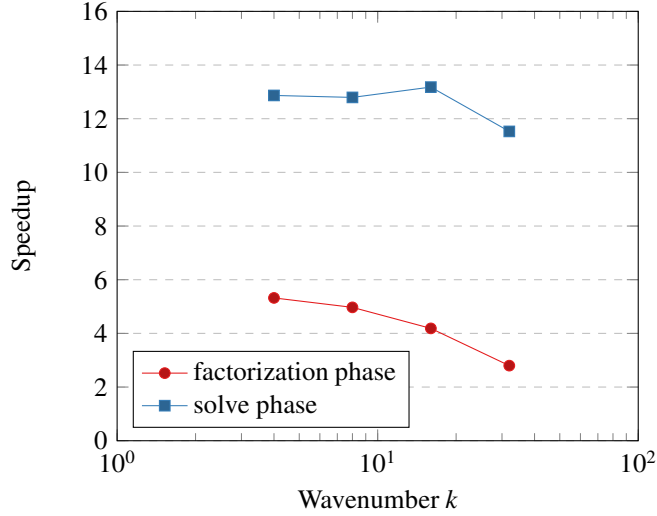}

  \fi

  \caption{Speedup of the node-based parallelization relative to the matrix-based parallelization in terms of the factorization and solve phases. Here, since the number of iterations (denoted by ``\niter'' in Tables II and III in the supplementary material) is preconditioner specific, the time of the solve phase was calculated by dividing the preconditioning time (which is spent for executing the preconditioning in \autoref{algo:precon}) by the number of applications (i.e., $\textrm{\niter}+2$).} 
  \label{fig:sphereii10_speedup}
\end{figure}

We compared the two parallel methods, using the best nIFMMsep3top3bot6 and mIFMMbot6 in \autoref{fig:sphereii10_speedup}. The factorization phase was actually boosted by the proposed node-based parallelization. The speedup became smaller with $k$. This is likely due to the fact that each operator becomes larger with $k$ and, thus, individual matrix operation can be performed more efficiently with multiple threads.

Regarding the solve phase, the node-based parallelization achieved about 12 times speedup. This significant speedup is due to the fact that the number of matrix operations involved in the solve phase is absolutely small and therefore the matrix-based parallelization has little chance to work efficiently. In addition, the parallelized solve algorithm is free from the serialization unlike the factorization phase. However, similarly to the factorization phase, the efficiency of the node-based parallelization tends to decrease with $k$.

\subsubsection{Influence of nIFMM's parameters}\label{s:sphereii_influence}

\autoref{tab:sphereii10_elim} gives a closer look at the elapsed times of the initialization and factorization phases level by level for the best nIFMM at $k=16$. We analyzed the influence of the parameters as follows:
\begin{enumerate}

\item Influence of $\ell$ on initialization:

To construct the extended matrix, the IFMM's initializer computes the FMM operators for every node and level (at line \ref{line:reuse} in \autoref{algo:fmbem}). If $\ell$ is chosen as $\lFMM$ ($=7$), the initializer can largely reuse the operators that FMM precomputed (at line \ref{line:precomp}), although reading/writing an operator from/to memory requires $O(p^4)$ cost; this is why the initializer spent about 100 seconds even when $\ell=7$. 

If we let $\ell<\lFMM$, we no longer initialize the levels from $\ell+1$ to $\lFMM$. On the other hand, we have to newly handle the (uncompressed) FMM operators at the leaf level $\ell$. The additional cost is estimated as $\sim 4^{-\ell}+4^\ell\ell^{-4}$ from the observations\footnote{The cost is proportional to the number of nodes and the square of the individual matrix size at the leaf level $\ell$. The former is $\sim 4^\ell$, while the latter is $\sim(4^{-\ell})^2$ for $\PtoP^{(\ell)}_{ij}$ and $\sim(p^2)^2\sim\ell^{-4}$ for others such as $\MtoL^{(\ell)}_{ij}$, where $p\sim\kappa^{-1}$ holds at level $\kappa$.} in \autoref{s:sphereii_best} and increases exponentially as $\ell$ decreases. The trade-off of the positive and negative effects can explain the trend of the initialization time in the table.

\item Influence of $\ell$ on factorization:

  Similarly to the above observation, if we let $\ell<\lFMM$, we can omit the factorization from level $\ell+1$ to $\lFMM$. On the other hand, the time can increase as $\ell$ decreases, because the cost at the leaf level $\ell$ can be estimated as $\sim 4^\ell \times (p^2)^3=4^\ell\ell^{-6}$ from some presumptions\footnote{The first factor $4^\ell$ denotes the number of nodes at leaf level $\ell$ and the second factor $(p^2)^3$ denotes the cost of a matrix-matrix product, which is the typical manipulation in the factorization, at the leaf level, where uncompressed FMM operators of size $\sim p^2\times p^2$ are manipulated}.  This cost function has the minimum value at $\ell=\frac{6}{\log 4}\approx 4.3$. This indicates that there exists an optimal value of $\ell$ that minimizes the cost at leaf level.

  In fact, \autoref{tab:sphereii10_elim} shows that, when we decreased $\ell$ from $7$ ($=\lFMM$) to 6, the time for the leaf level almost remained, but the total time was decreased by the time for level 7, i.e., about 47 seconds. This is the positive effect of decreasing $\ell$. On the other hand, when we decreased $\ell$ from 6 to 5 or 4, both time for the leaf level and total time increased significantly. This corresponds to the negative effect and means that the minimizer was actually around 6 rather than 4.3.

\item Influence of $\tau$ on factorization:
  
  In \autoref{tab:sphereii10_elim}, $\tau=3$ has no clear difference from $\tau=2$ at level 2, while $\tau=4$ certainly spoiled the performance at level 3, to which the matrix-based parallelization is applied. This is because the size of each matrix (operator) at level 3 is too small to cultivate an enough parallel efficiency with 16 computing cores.
  
\end{enumerate}

\begin{table}[h]
  \centering
  \caption{Breakdown of the factorization and initialization times (in seconds) of nIFMM in the sphere model.} 
  \label{tab:sphereii10_elim}
  \def\mydata{./181031/sphereii}
  $\varepsilon=10^{-3}$, $k=16$\\
  \begin{tabular}{|l|r|r|r|r|r|r|r|r|} 
    \toprule
    Precon. & \multicolumn{7}{c|}{Factorization} & Initialization\\
            & Total &\multicolumn{6}{c|}{Level} & Total\\
            &       & 2 & 3 & 4 & 5 & 6 & 7 &\\
    \midrule
    nIFMMs3t2b7 & \prtime{ 124.886904} & \prtime{5.64113} & \prtime{5.62803} & \prtime{8.51921} & \prtime{16.8236} & \prtime{39.9745} & \prtime{46.8716}&\prtime{ 99.531408}\\

    nIFMMs3t2b6 & \prtime{ 81.648971} & \prtime{5.33013} & \prtime{6.17701} & \prtime{7.95789} & \prtime{16.1394} & \prtime{45.6436}& &\prtime{ 101.304549}\\

    nIFMMs3t2b5 & \prtime{ 122.251087} & \prtime{4.0602} & \prtime{5.65707} & \prtime{9.42954} & \prtime{102.99858}& & &\prtime{ 435.314870}\\

    nIFMMs3t2b4 & \prtime{ 984.549876} & \prtime{4.24479} & \prtime{6.2789} & \prtime{974.0007}& & & &\prtime{ 1843.679789}\\

    \midrule
    nIFMMs3t3b7 & \prtime{ 128.582756} & \prtime{7.48356} & \prtime{5.70098} & \prtime{8.64864} & \prtime{17.0025} & \prtime{39.9656} & \prtime{48.2873}&\prtime{ 102.818790}\\

    nIFMMs3t3b6 & \prtime{ 81.245963} & \prtime{5.36098} & \prtime{6.12108} & \prtime{7.91494} & \prtime{16.3276} & \prtime{45.1346}& &\prtime{ 101.911095}\\

    nIFMMs3t3b5 & \prtime{ 121.653466} & \prtime{3.41884} & \prtime{5.6525} & \prtime{9.44492} & \prtime{103.03944}& & &\prtime{ 418.787389}\\

    nIFMMs3t3b4 & \prtime{ 978.992829} & \prtime{4.63972} & \prtime{6.24653} & \prtime{968.0817}& & & &\prtime{ 1802.981907}\\

    \midrule
    nIFMMs3t4b7 & \prtime{ 131.998320} & \prtime{5.89141} & \prtime{11.1767} & \prtime{8.60473} & \prtime{16.8103} & \prtime{39.919} & \prtime{48.1031}&\prtime{ 103.658072}\\

    nIFMMs3t4b6 & \prtime{ 85.982191} & \prtime{5.06449} & \prtime{11.1924} & \prtime{7.89645} & \prtime{16.0921} & \prtime{45.3360}& &\prtime{ 101.699027}\\

    nIFMMs3t4b5 & \prtime{ 126.295347} & \prtime{2.80649} & \prtime{10.0327} & \prtime{9.51767} & \prtime{103.83039}& & &\prtime{ 413.236559}\\

    nIFMMs3t4b4 & \prtime{ 987.142215} & \prtime{4.63504} & \prtime{11.493} & \prtime{970.9873}& & & &\prtime{ 1756.588535}\\

    \bottomrule
  \end{tabular}\\
\end{table}


\subsubsection{Basic parallel algorithm ($\separation=6$) vs improved parallel  algorithm ($\separation=3$)}\label{s:sphereii_sep6}

We compared the improved parallel algorithm based on $\separation=3$ with the basic parallel algorithm based on $\separation=6$. The latter is free from any serialized operations, but the group size is about eight ($=6^3/3^3$) times smaller than $\separation=3$ in average.

When we varied $\tau$ from $2$ to $4$ and $\ell$ from $4$ to $7$ in the case of $\separation=6$, the best performance was obtained by nIFMMsep6top4bot6, although the value of $\tau$ was almost insensitive to the timing result.

\begin{figure}[h]
  \centering

  \iftrue 
  
  
  \begin{tikzpicture}
    \begin{axis}[
      xlabel={Wavenumber $k$},
      ylabel={Speedup of $\separation=3$ relative to $\separation=6$},
      xmode=log,
      xmin=1, xmax=100,
      ymin=1.0, ymax=2.5, ytick={1.0, 1.5,..., 2.5},
      legend pos=north west,
      legend cell align={left},
      ymajorgrids=true,
      grid style=dashed,
      legend entries={factorization phase, solve phase, total}
      ]

      \addplot table [x=k, y=sp_ifmm_elim] {./181031/sphereii/fig-nIFMM_relto_nIFMM6-sphereii10-delta7-epsilon1e-3.table}; 
      \addplot table [x=k, y=sp_msolve_per_iter] {./181031/sphereii//fig-nIFMM_relto_nIFMM6-sphereii10-delta7-epsilon1e-3.table}; 
      \addplot table [x=k, y=sp_main] {./181031/sphereii//fig-nIFMM_relto_nIFMM6-sphereii10-delta7-epsilon1e-3.table}; 
      
    \end{axis}
  \end{tikzpicture}

  \else

  \begin{tabular}{c}
    \includegraphics[width=.5\textwidth]{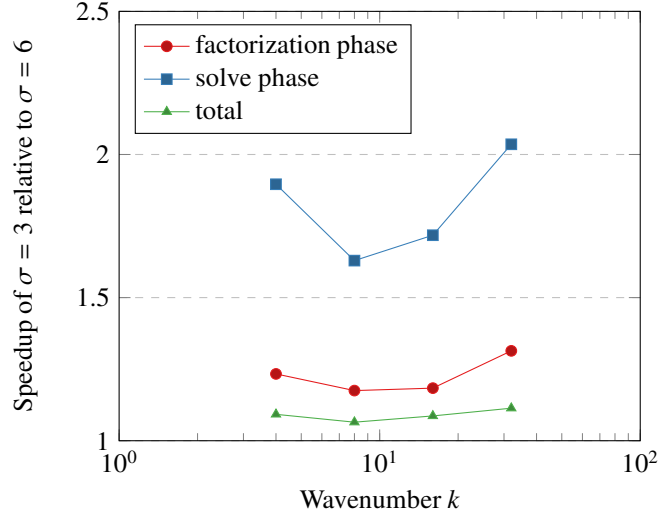}
  \end{tabular}

  \fi

  \caption{Speedup of the best nIFMM based on $\separation=\textbf{3}$ (i.e., nIFMM\textbf{sep3}top3bot6) relative to the best one based on $\separation=\textbf{6}$ (i.e., nIFMM\textbf{sep6}top4bot6) in the sphere model. Here, the time of the solve phase was calculated by dividing the preconditioning time by $\textrm{\niter}+2$; see the caption of \autoref{fig:sphereii10_speedup}.} 

  \label{fig:sphereii10_sep6}
\end{figure}

\autoref{fig:sphereii10_sep6} compares nIFMMsep6top4bot6 with nIFMMsep3top3bot6 (i.e., the best nIFMM using $\separation=3$). Although the difference in total computation time is not significant, this figure shows that $\separation=3$ was indeed better than $\separation=6$ for any $k$.

\subsubsection{Influence of $\varepsilon$ and accuracy}\label{s:sphereii_accuracy}

We tested relatively low-precision $\varepsilon$, i.e., $\varepsilon=10^{-1}$ and $10^{-2}$, for mIFMMbot6 and nIFMMsep3top3bot6, which are the best in the case of $\varepsilon=10^{-3}$. A larger $\varepsilon$ led to a shorter initialization and factorization time but resulted in a larger number of iterations. As a result, $\varepsilon=10^{-2}$ maximized the performances of both mIFMMbot6 and nIFMMsep3top3bot6 in the sphere model.

In all the above analyses, the accuracy of the solutions obtained by all the preconditioners were almost the same for every $k$. In particular, $\varepsilon$ did not affect the accuracy; this is shown in Tables~I, II, and III in the supplementary material.

\subsubsection{Memory consumption}

Regarding the largest $k=32$, nIFMMsep3top3bot6 required 381, 393, and \unit{410}{GB} for $\varepsilon=10^{-1}$, $10^{-2}$, and $10^{-3}$, respectively. About 90\% of the required memory was spent for constructing the extended matrix $\bar{A}$ in the initialization phase. Since this phase is common, mIFMMbot6 required almost the same memory as the nIFMM. The huge memory consumption is problematic, but will be less of an issue with distributed memory.

On the other hand, BDbot5 required \unit{25}{GB} for $k=32$, where the precomputation of M2L, P2P, and the preconditioning matrix $M$ spent 9, 6, and \unit{5}{GB}, respectively.


\subsection{Realistic example --- office model}\label{s:office}

\subsubsection{Problem statement}

\begin{figure}[h]
  \includegraphics[width=.25\textwidth]{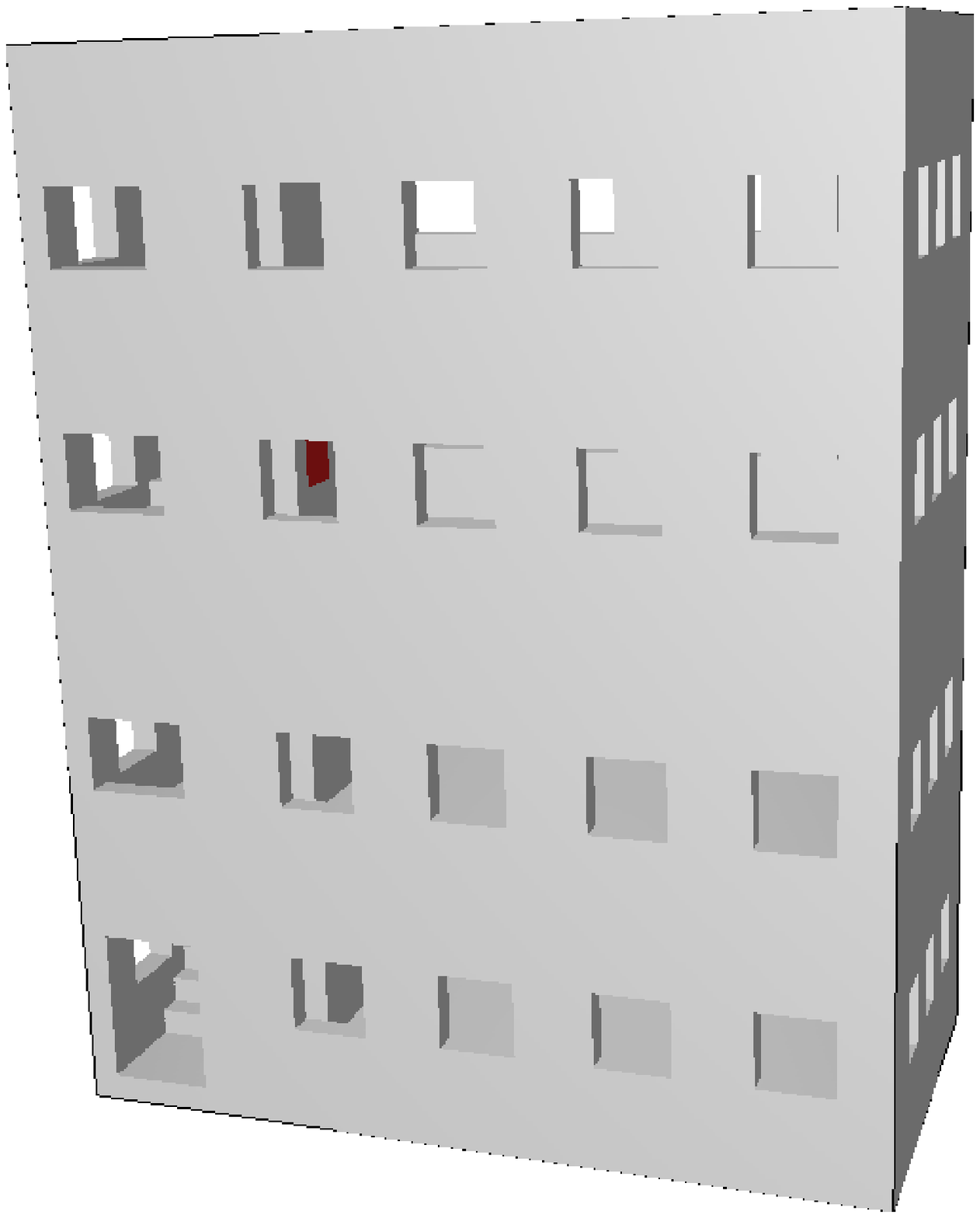} 
  \quad
  \includegraphics[width=.25\textwidth]{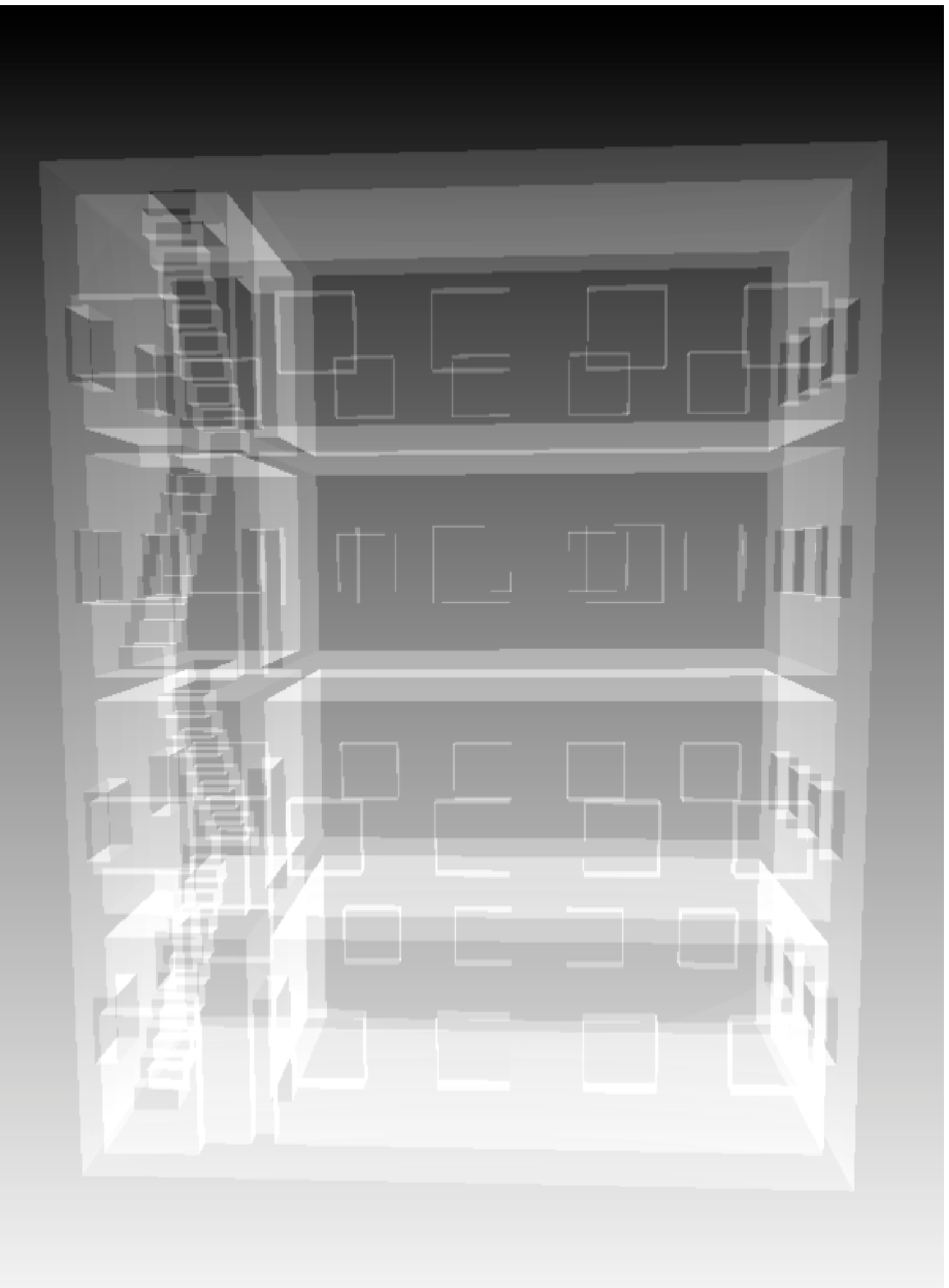}
  \quad
  \includegraphics[width=.4\textwidth]{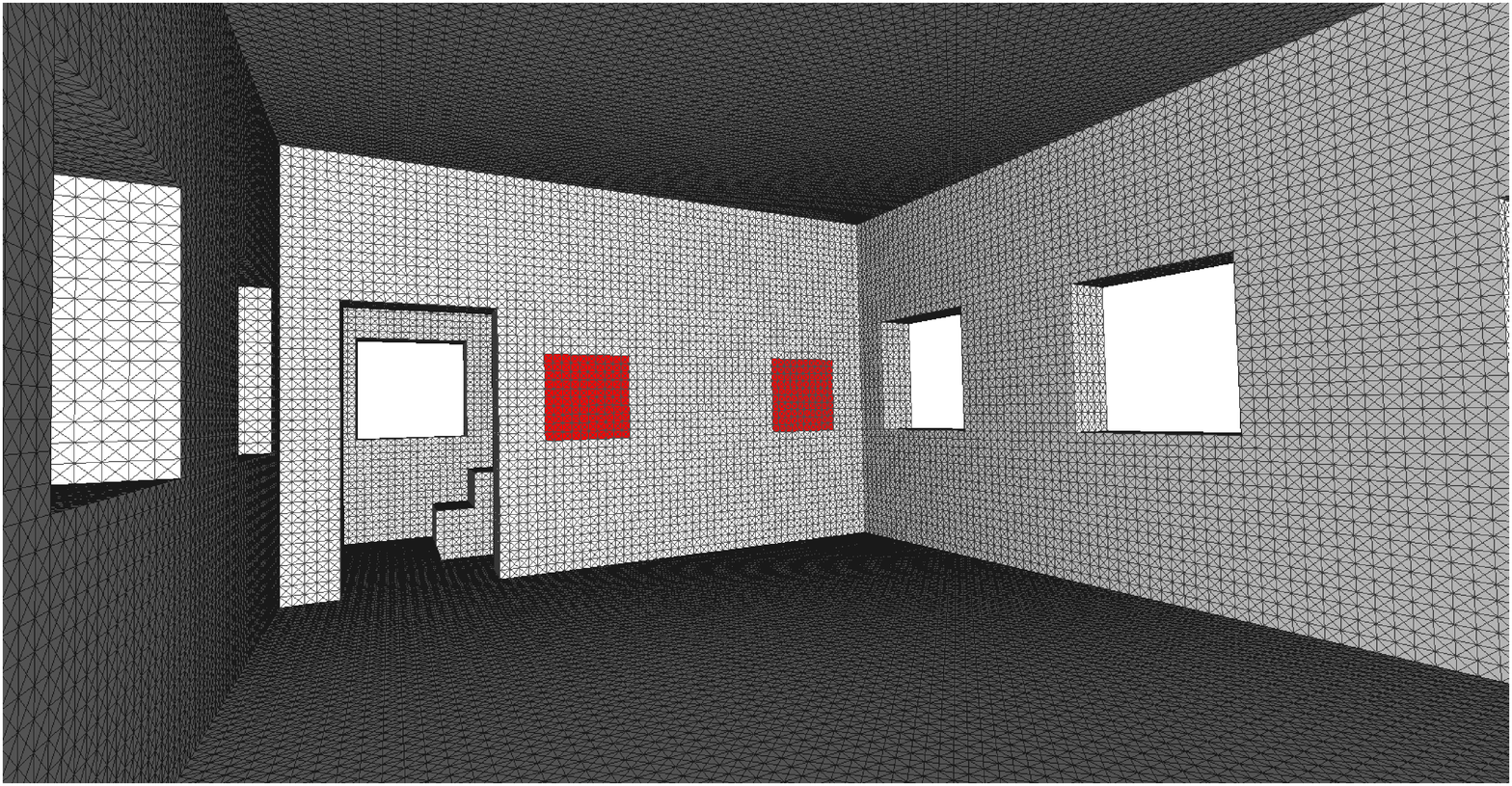}
  \caption{Office model. (Left) Outlook. The dimensions $\text{width}\times\text{depth}\times\text{height}$ are $0.8\times 0.4 \times 1.025$. (Center) Transparent view. (Right) Two sound sources (colored in red) on the wall in the third floor and the boundary element mesh.}
  \label{fig:model}
\end{figure}

We consider a complicated problem in comparison with the previous one. Namely, we solve a mixed boundary value problem (BVP) where the scatterer is an office building consisting of four floors (\autoref{fig:model} left and center). Regarding the boundary condition, the Dirichlet boundary condition of $u=1$ is given as the sound source on a wall in the third floor (\autoref{fig:model} right), while the Neumann boundary condition of $q=0$ is given everywhere else. The model is discretized with 1218200 elements.

We chose the maximum number of elements per node ($\mu$) as $40$, following the previous sphere model. Then, the depth of the FMM hierarchy (i.e., $\lFMM$) became $7$ as shown in \autoref{tab:stat_office05}.

\begin{table}[h] 
  \def\prave#1{{\nprounddigits{0}\numprint{#1}}}
  \centering
  \caption{Statistics for the \textbf{office} model. The meanings of the headings are the same as those in \autoref{tab:stat_sphereii10}.}
  \label{tab:stat_office05}
  %
  \small
  \begin{tabular}{|c|r|r|r|r|r|r|r|r|r|}
    \toprule
    Level  & \multicolumn{2}{c|}{FMM} & IFMM & \multicolumn{4}{c|}{nIFMM} & \multicolumn{2}{c|}{$p^2$}\\
    $\kappa$ & \#nodes & \#ele/node & \#nodes & \#groups & Min & Ave & Max & $k=4$ & $k=32$\\
    \midrule
    2 & 32 & \prave{38068.750000} & 32 & 18 & 1 & \prave{1.777778} & 4 & 144 & 2704\\
    3 & 256 & \prave{4758.593750} & 256 & 27 & 4 & \prave{9.481481} & 18 & 100 & 1156\\
    4 & 1512 & \prave{805.687831} & 1512 & 27 & 10 & \prave{56.000000} & 86 & 81 & 529\\
    5 & 6775 & \prave{179.808118} & 6775 & 27 & 76 & \prave{250.92592} & 347 & 64 & 256\\
    6 & 28867 & \prave{42.186822} & 28890 & 26 & 116 & \prave{1111.153846} & 1512 & 64 & 144\\
    7 & 65264 & \prave{11.295477} & 115789 & 19 & 4002 & \prave{6094.157895} & 9089 & 64 & 100\\
    \bottomrule
  \end{tabular}
\end{table}

\subsubsection{Comparison under the best settings}\label{s:office_comp3}

We focus on our prime interest, that is, the performance of nIFMM over BD and mIFMM. Prior to this, we surveyed the best setting of each preconditioner; the details are described in Section~2 of the supplementary material. Basically, since the present model is more complicated, the number of iterations became larger than the previous model. Hence, a longer initialization time can compensate the total computation time. We actually considered $\lBD\in\{2,7\}$, $\separation=3$, $\tau\in\{2,3\}$, and $\ell\in\{4,\ldots,7\}$ for the survey.

As a result, BDbot4 was nearly the best for $k\le 32$ and BDbot3 for $k=64$. Meanwhile, mIFMMbot5 and nIFMMsep3top3bot5 were nearly the best for higher wavenumbers. 

\begin{figure}[h]
  \centering

  \iftrue 

  \begin{tikzpicture}
    \begin{axis}[
      xlabel={Wavenumber $k$},
      ylabel style={align=center},
      ylabel={Speedup of nIFMMs3t3b5\\relative to BD and mIFMMb5},
      xmode=log, ymode=log,
      xmin=1, xmax=100,
      ymin=0.1, ymax=100.0,
      legend pos=north east,
      legend cell align={left},
      ymajorgrids=true,
      grid style=dashed,
      legend entries={BD, mIFMMb5},
      ]
      
      \addplot table [x=k, y=tsp_nIFMM] {./181031/office/fig-tsp-nIFMM-office05-delta7-epsilon1e-3.table};
      \addplot table [x=k, y expr=\thisrow{tsp_nIFMM}/\thisrow{tsp_mIFMM}] {./181031/office/fig-tsp-nIFMM-office05-delta7-epsilon1e-3.table};
      
    \end{axis}
  \end{tikzpicture}

  \else

  \includegraphics[width=.5\textwidth]{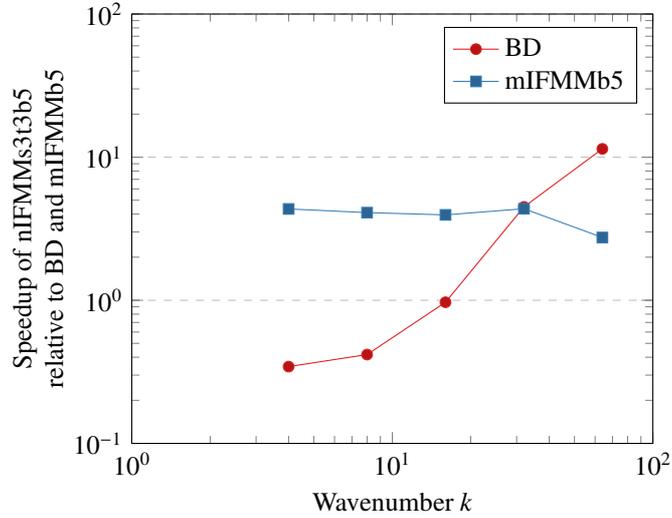}

  \fi

  \caption{Speedup off the best nIFMM (i.e., nIFMMsep3top3bot5) relative to the best mIFMM (i.e., mIFMMbot5) and BD in terms of the total computation time for the \textbf{office} model.} 
  \label{fig:office05_speedup}
\end{figure}

\autoref{fig:office05_speedup} shows the speedup of nIFMMsep3top3bot5 relative to mIFMMbot5 and BD in terms of the total computation time. Here, the time of BD was chosen as the best (shortest) time of BDbot3--7 for every wavenumber; this is because it is difficult to determine the best BD for all the wavenumbers. Clearly, the nIFMM increased its performance with $k$ and achieved about {\nprounddigits{1}\numprint{11.42854921289711542379}} times speedup at the largest wavenumber. Meanwhile, the speedup over the mIFMM was about 4 times.

\subsubsection{Basic parallel algorithm ($\separation=6$) vs improved parallel algorithm ($\separation=3$)}\label{s:office_sep6}

\begin{figure}[h]
  \centering

  \iftrue 

  \begin{tikzpicture}
    \begin{axis}[
      xlabel={Wavenumber $k$},
      ylabel={Speedup of $\separation=3$ relative to $\separation=6$},
      xmode=log,
      xmin=1, xmax=100,
      ymin=1.0, ymax=3.0, ytick={1.0, 1.5,..., 3.0},
      legend pos=north west,
      legend cell align={left},
      ymajorgrids=true,
      grid style=dashed,
      legend entries={factorization phase, solve phase, total}
      ]
      
      \addplot table [x=k, y=sp_ifmm_elim] {./181031/office/fig-nIFMM_relto_nIFMM6-office05-delta7-epsilon1e-3.table}; 
      \addplot table [x=k, y=sp_msolve_per_iter] {./181031/office/fig-nIFMM_relto_nIFMM6-office05-delta7-epsilon1e-3.table}; 
      \addplot table [x=k, y=sp_main] {./181031/office/fig-nIFMM_relto_nIFMM6-office05-delta7-epsilon1e-3.table}; 
      
    \end{axis}
  \end{tikzpicture}

  \else
 
  \begin{tabular}{c}
    \includegraphics[width=.5\textwidth]{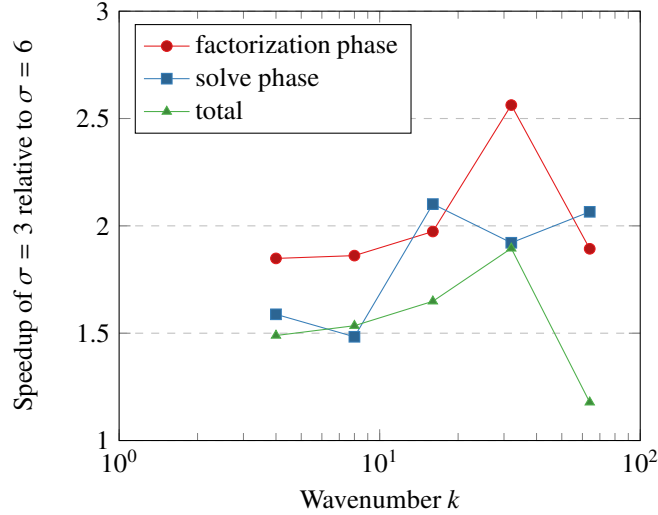}
  \end{tabular}

  \fi
 
  \caption{Speedup of nIFMM\textbf{sep3}top3bot5 relative to nIFMM\textbf{sep6}top4bot5 in the \textbf{office} model. The meanings of the legends are the same as those in \autoref{fig:sphereii10_sep6}.} 
  \label{fig:office05_sep6}
\end{figure}

We simply changed the value of $\separation$ from $3$ to $6$ in the best nIFMM (i.e., nIFMMsep3top3bot5). \autoref{fig:office05_sep6} shows that $\separation=3$ is indeed faster than $\separation=6$ for any wavenumbers. Similarly to the previous example in \autoref{s:sphereii_sep6}, we can confirm that the serialization of a small fraction of the factorization algorithm pays the total efficiency.

\subsubsection{Accuracy}\label{s:office_accuracy}

Finally, we point out that the IFMM-based preconditioners with using larger values of $\varepsilon$, that is, $10^{-1}$ and $10^{-2}$, produced inaccurate solutions in some cases, where the distribution of the sound pressure $u$ on the boundary were erroneous at specific portions.

The previous work~\cite{takahashi2017} suggested that $\varepsilon=10^{-2}$ was a safer choice by considering the fact that the erroneous solutions were induced by $\varepsilon=10^{-1}$ in a similar mixed BVP of $N=147$k and $k\le 32$. However, this is not true for the present problem of $N=1218$k and $k\le 64$. The present experiments conclude $\varepsilon=10^{-3}$ is safer. In addition, $\varepsilon=10^{-3}$ was faster than $\varepsilon=10^{-1}$ and $10^{-2}$ for most of cases of $k$ because $\varepsilon=10^{-3}$ resulted in more rapid convergence. 

Choosing an optimal value of the IFMM's precision parameter $\varepsilon$ still remains an open question regardless of the parallelization.

\section{Conclusion}\label{s:concl}

To perform IFMM (\autoref{s:ifmm}) efficiently, we proposed to parallelize its factorization and solve phases with respect to nodes in every level of the hierarchy. Unlike in FMM, a loop over nodes cannot be parallelized straightforwardly due to the data dependencies among nodes. To overcome it, we divide nodes in a level into groups (colors) so that any two nodes in each group are separated by $\separation L$ or more, where $\separation$ is a predefined parameter (termed to separation parameter) and $L$ denotes the edge length of the nodes (\autoref{s:grouping}). The number of groups is bounded from above by $\separation^3$. Then, if $\separation$ is 6 or more, we can prove that all the nodes (or computing threads assigned to them) in a group never result in any race condition and, thus, can be handled concurrently (\autoref{s:race}). The grouping (coloring) with $\separation=6$ is our basic strategy (\autoref{s:full}). However, the resulting number of nodes per group is not so large especially for upper levels. The number $\separation$ should be large for a high parallel efficiency. We thus considered to use a smaller value of $\separation$ and chose the minimum value of 3 (\autoref{s:part}). This can increase the number of nodes per group at most $8$ ($=6^3/3^3$) times. In this case, we can still parallelize the loop over nodes by serializing a small fraction of the factorization. We implemented a parallel IFMM code with OpenMP, which can run on a shared memory machine.

In the later part of this paper, we applied the parallel IFMM to a fast iterative BEM, which can be used for 3D acoustic scattering problems, as the preconditioner (\autoref{s:preconditioner}). The numerical results in \autoref{s:num} confirmed that the preconditioner based on the proposed ``n''ode-based parallelization (termed to \textbf{nIFMM}) was indeed faster than the ``m''atrix-based parallelization (termed to \textbf{mIFMM}). Also, nIFMM outperformed the extended block-diagonal preconditioner (termed to \textbf{BD}) for relatively large wavenumbers. Here, the numbers of unknowns are about one million and the size of scatterer is up to ten times wavelength.

In these performance comparisons, we optimized the bottom level (depth) of the hierarchy (i.e., $\ell$ for both nIFMM and mIFMM and $\lBD$ for BD), which can be chosen differently from the FMM's hierarchy (i.e., $\lFMM$). We discussed the tendency of the timing result due to $\ell$, but a way to predetermine an optimal choice has not been established yet. Regarding nIFMM, we also investigated the separation parameter $\separation$ and the top level $\tau$ as the tunable parameters. We found that $\tau$ was not so influential to the timing result, while we observed that $\separation=3$ was indeed better than $\separation=6$.

Finally, we remark the future directions of this study:
\begin{itemize}

\item First, in order to make full use of threads, we will realize the nested parallelization, that is, parallelize each matrix operation in the parallelized loop over nodes. This is not allowed by the current version of Eigen~\cite{eigenweb}, which is used as the front-end of the MKL.  This can be regarded as a hybrid of the node- and matrix-based parallelizations, which are alternative in a level in this study. Such a fine (and complicated) control of threads would be possible by utilizing both Intel MKL~\cite{Intel_MKL} and OpenMP~\cite{OpenMP} (without Eigen) but technically challenging.

\item Second, we will implement a parallel IFMM code for memory-distributed systems, using a message passing interface (MPI). This parallelization could resolve the issue of huge memory consumption, which we experienced on shared-memory systems. For this regard, the parallelization proposed for a sparse direct solver (LoRaSp~\cite{pouransari2017}) by Chen et al~\cite{chen2018} would be helpful. However, the modification to dense systems is not trivial because the algorithm of IFMM is significantly different from that of LoRaSp. Once this parallelization using MPI is done, it could be combined with the proposed (or the aforementioned nested) parallelization using OpenMP to construct a so-called MPI-OpenMP hybrid parallelization, which is a good fit for modern cluster computing systems.

\end{itemize}

\section*{Acknowledgments}\label{s:acks}
The work of the first author is partially supported by JSPS KAKENHI under Grant Number 18K11335.


Part of this research was done at Stanford University. This material is based upon the work supported by the Department of Energy under Award Number DE-NA0002373-1.

Some of the computing for this project was performed on the Sherlock cluster 2.0. We would like to thank Stanford University and the Stanford Research Computing Center for providing computational resources and support that contributed to these research results.


\bibliographystyle{elsarticle/elsarticle-num}
\bibliography{paper}

\begin{thebibliography}{10}
\expandafter\ifx\csname url\endcsname\relax
  \def\url#1{\texttt{#1}}\fi
\expandafter\ifx\csname urlprefix\endcsname\relax\def\urlprefix{URL }\fi
\expandafter\ifx\csname href\endcsname\relax
  \def\href#1#2{#2} \def\path#1{#1}\fi

\bibitem{1985_Rokhlin}
V.~Rokhlin, Rapid solution of integral equations of classical potential theory,
  Journal of Computational Physics 60~(2) (1985) 187--207.
\newblock

\bibitem{1987_Greengard}
L.~Greengard, V.~Rokhlin, A fast algorithm for particle simulations, Journal of
  Computational Physics 73~(2) (1987) 325--348.
\newblock

\bibitem{Darve2000195}
E.~Darve, The fast multipole method: Numerical implementation, Journal of
  Computational Physics 160~(1) (2000) 195--240.
\newblock

\bibitem{2002_Nishimura}
N.~Nishimura, Fast multipole accelerated boundary integral equation methods,
  Applied Mechanics Reviews 55~(4) (2002) 299--324.
\newblock

\bibitem{2014_Ambikasaran}
S.~{Ambikasaran}, E.~{Darve}, The inverse fast multipole method, ArXiv
  e-prints\href {http://arxiv.org/abs/1407.1572} {\path{arXiv:1407.1572}}.

\bibitem{coulier2017}
P.~Coulier, H.~Pouransari, E.~Darve, The inverse fast multipole method: Using a
  fast approximate direct solver as a preconditioner for dense linear systems,
  SIAM Journal on Scientific Computing 39~(3) (2017) A761--A796.

\bibitem{2005_Martinsson}
P.~Martinsson, V.~Rokhlin, A fast direct solver for boundary integral equations
  in two dimensions, Journal of Computational Physics 205~(1) (2005) 1--23.
\newblock

\bibitem{2015_Corona}
E.~Corona, P.-G. Martinsson, D.~Zorin, An direct solver for integral equations
  on the plane, Applied and Computational Harmonic Analysis 38~(2) (2015)
  284--317.

\bibitem{2015_Bremer}
J.~Bremer, A.~Gillman, P.-G. Martinsson, A high-order accurate accelerated
  direct solver for acoustic scattering from surfaces, BIT Numerical
  Mathematics 55~(2) (2015) 367--397.

\bibitem{2016_Ho}
K.~L. Ho, L.~Ying, Hierarchical interpolative factorization for elliptic
  operators: Differential equations, Communications on Pure and Applied
  Mathematics 69~(8) (2016) 1415--1451.

\bibitem{bebendorf2008hierarchical}
M.~Bebendorf, Hierarchical Matrices: A Means to Efficiently Solve Elliptic
  Boundary Value Problems, Lecture Notes in Computational Science and
  Engineering, Springer Berlin Heidelberg, 2008.

\bibitem{2008_Banjai}
L.~Banjai, W.~Hackbusch, {Hierarchical matrix techniques for low- and
  high-frequency {Helmholtz} problems}, {IMA Journal of Numerical Analysis}
  28~(1) (2008) 46--79.
\newblock

\bibitem{quaife2017}
B.~Quaife, P.~Coulier, E.~Darve,
  \href{https://onlinelibrary.wiley.com/doi/abs/10.1002/nme.5626}{An efficient
  preconditioner for the fast simulation of a {2D} {Stokes} flow in porous
  media}, International Journal for Numerical Methods in Engineering 113~(4)
  (2018) 561--580.
\newblock
\newline\urlprefix\url{https://onlinelibrary.wiley.com/doi/abs/10.1002/nme.5626}

\bibitem{saad1986gmres}
Y.~Saad, M.~H. Schultz, Gmres: A generalized minimal residual algorithm for
  solving nonsymmetric linear systems, SIAM Journal on scientific and
  statistical computing 7~(3) (1986) 856--869.

\bibitem{takahashi2017}
T.~Takahashi, P.~Coulier, E.~Darve,
  \href{http://www.sciencedirect.com/science/article/pii/S0021999117302875}{Application
  of the inverse fast multipole method as a preconditioner in a {3D}
  {Helmholtz} boundary element method}, Journal of Computational Physics 341
  (2017) 406--428.
\newblock
\newline\urlprefix\url{http://www.sciencedirect.com/science/article/pii/S0021999117302875}

\bibitem{ij-cmame-coul-16a}
P.~Coulier, E.~Darve, Efficient mesh deformation based on radial basis function
  interpolation by means of the inverse fast multipole method, Computer Methods
  in Applied Mechanics and Engineering 308 (2016) 286--309.
\newblock

\bibitem{2011_Darve}
E.~Darve, C.~Cecka, T.~Takahashi, The fast multipole method on parallel
  clusters, multicore processors, and graphics processing units, Comptes Rendus
  Mécanique 339~(2-3) (2011) 185--193.
\newblock

\bibitem{yokota2011biomolecular}
R.~Yokota, J.~P. Bardhan, M.~G. Knepley, L.~Barba, T.~Hamada, Biomolecular
  electrostatics using a fast multipole {BEM} on up to 512 {GPU}s and a billion
  unknowns, Computer Physics Communications 182~(6) (2011) 1272 -- 1283.

\bibitem{saad1996iterative}
Y.~Saad, Iterative methods for sparse linear systems, pws pub, Co., Boston.

\bibitem{wang2014intel}
E.~Wang, Q.~Zhang, B.~Shen, G.~Zhang, X.~Lu, Q.~Wu, Y.~Wang, Intel math kernel
  library, in: High-Performance Computing on the Intel{\textregistered} Xeon
  Phi\texttrademark, Springer, 2014, pp. 167--188.

\bibitem{OpenBLAS}
\href{https://github.com/xianyi/OpenBLAS}{{OpenBLAS}}.
\newline\urlprefix\url{https://github.com/xianyi/OpenBLAS}

\bibitem{chen2018}
C.~Chen, H.~Pouransari, S.~Rajamanickam, E.~G. Boman, E.~Darve,
  \href{http://www.sciencedirect.com/science/article/pii/S0167819117302077}{A
  distributed-memory hierarchical solver for general sparse linear systems},
  Parallel Computing 74 (2018) 49--64.
\newblock
\newline\urlprefix\url{http://www.sciencedirect.com/science/article/pii/S0167819117302077}

\bibitem{pouransari2017}
H.~Pouransari, P.~Coulier, E.~Darve,
  \href{https://doi.org/10.1137/15M1046939}{Fast hierarchical solvers for
  sparse matrices using extended sparsification and low-rank approximation},
  SIAM Journal on Scientific Computing 39~(3) (2017) A797--A830.
\newblock
\newline\urlprefix\url{https://doi.org/10.1137/15M1046939}

\bibitem{OpenMP}
\href{http://openmp.org/wp}{{The OpenMP API Specification for Parallel
  Programming}}.
\newline\urlprefix\url{http://openmp.org/wp}

\bibitem{ZoltanIsorropiaOverview2012}
E.~G. Boman, U.~V. Catalyurek, C.~Chevalier, K.~D. Devine, The {Z}oltan and
  {I}sorropia parallel toolkits for combinatorial scientific computing:
  Partitioning, ordering, and coloring, Scientific Programming 20~(2) (2012)
  129--150.

\bibitem{1971_Burton_Miller}
A.~J. Burton, G.~F. Miller, The application of integral equation methods to the
  numerical solution of some exterior boundary-value problems, Proceedings of
  the Royal Society of London A: Mathematical, Physical and Engineering
  Sciences 323~(1553) (1971) 201--210.
\newblock

\bibitem{2003_Saad_book}
Y.~Saad, Iterative Methods for Sparse Linear Systems, 2nd Edition, Society for
  Industrial and Applied Mathematics, Philadelphia, PA, USA, 2003.

\bibitem{1995_Epton}
M.~A. Epton, B.~Dembart, Multipole translation theory for the three-dimensional
  {Laplace} and {Helmholtz} equations, SIAM J. Sci. Comput. 16~(4) (1995)
  865--897.
\newblock

\bibitem{2009_Liu_book}
Y.~Liu, Fast Multipole Boundary Element Method: Theory and Applications in
  Engineering, Cambridge University Press, Cambridge, 2009.

\bibitem{2011_Liu}
Y.~J. Liu, S.~Mukherjee, N.~Nishimura, M.~Schanz, W.~Ye, A.~Sutradhar, E.~Pan,
  N.~A. Dumont, A.~Frangi, A.~Saez, Recent advances and emerging applications
  of the boundary element method, Applied Mechanics Reviews 64~(3) (2011)
  030802.
\newblock

\bibitem{2016_Takahashi}
T.~Takahashi, Y.~Shimba, H.~Isakari, T.~Matsumoto, An efficient blocking {M2L}
  translation for low-frequency fast multipole method in three dimensions,
  Computer Physics Communications 202 (2016) 151--164.
\newblock

\bibitem{darv04b}
E.~Darve, P.~Hav\'e, Efficient fast multipole method for low-frequency
  scattering, Journal of Computational Physics 197~(1) (2004) 341--363.

\bibitem{Intel_MKL}
\href{https://software.intel.com/en-us/intel-mkl}{{Intel Math Kernel Library
  (Intel MKL)}}.
\newline\urlprefix\url{https://software.intel.com/en-us/intel-mkl}

\bibitem{eigenweb}
G.~Guennebaud, B.~Jacob, et~al., {Eigen v3}, http://eigen.tuxfamily.org (2010).

\bibitem{fong2007}
W.~Fong, E.~Darve, The black-box fast multipole method, Journal of
  Computational Physics 228 (2009) 8712--8725.
\newblock

\bibitem{1993_Coifman}
R.~Coifman, V.~Rokhlin, S.~Wandzura, {The fast multipole method for the wave
  equation: A pedestrian prescription}, Antennas and Propagation Magazine, IEEE
  35~(3) (1993) 7--12.
\newblock

\bibitem{2011_Halko}
N.~Halko, P.~G. Martinsson, J.~A. Tropp, Finding structure with randomness:
  Probabilistic algorithms for constructing approximate matrix decompositions,
  SIAM Review 53~(2) (2011) 217--288.
\newblock

\bibitem{Bowman}
J.~J. Bowman, T.~B.~A. Senior, P.~L.~E. Uslenghi, {Electromagnetic and Acoustic
  Scattering by Simple Shapes (Revised edition)}, Hemisphere Publishing Corp.,
  New York, 1987.

\end{thebibliography}

\end{document}